\documentclass{article}
\usepackage{epsfig}
\usepackage{physics_of_fluids}
\begin{document}

\begin{titlepage}

\title{On acoustic cavitation of slightly subcritical bubbles}

\author{Anthony Harkin\raisebox{0.1in}{$\dagger$}
\hspace{0.2in}
Ali Nadim\raisebox{0.1in}{$\ddagger$}
\hspace{0.2in}
Tasso J. Kaper\raisebox{0.1in}{$\dagger$}
\\
\footnotesize{\raisebox{0.075in}{$\dagger$}Department of Mathematics,
Boston University, Boston, MA 02215}
\\
\footnotesize{\raisebox{0.075in}{$\ddagger$}Department of Aerospace
and Mechanical Engineering,
Boston University, Boston, MA 02215}
}

\date{July 24, 1998}

\maketitle

\begin{abstract}
The classical Blake threshold indicates the onset of quasistatic
evolution leading to cavitation for gas bubbles in liquids. When the
mean pressure in the liquid is reduced to a value below the vapor
pressure, the Blake analysis identifies a critical radius which
separates quasistatically stable bubbles from those which would
cavitate. In this work, we analyze the cavitation threshold for
radially symmetric bubbles whose radii are slightly less than the
Blake critical radius, in the presence of time-periodic
acoustic pressure fields.  A distinguished limit equation is derived
that predicts the threshold for cavitation for a wide range of liquid
viscosities and forcing frequencies.  This equation also yields
frequency-amplitude response curves.  Moreover, for fixed liquid
viscosity, our study identifies the frequency that yields the minimal
forcing amplitude sufficient to initiate cavitation.  Numerical
simulations of the full Rayleigh-Plesset equation confirm the accuracy
of these predictions.  Finally, the implications of these findings for 
acoustic pressure fields that consist of two frequencies will be discussed.
\end{abstract}

\bigskip
\noindent
{\bf PACS Numbers:}
Primary 43.25.Yw,
Secondary 43.25.Ts, 47.52.+j, 43.25.Rq

\bigskip
\noindent
{\bf Keywords:}
acoustic cavitation, nonlinear oscillations of gas bubbles,
dynamic cavitation threshold, periodic pressure fields,
quasiperiodic pressure fields, period-doubling.

\thispagestyle{empty}
\end{titlepage}

\addtocounter{page}{1}

\section{Introduction}
\label{Introduction}

The Blake threshold pressure is the standard measure of static
acoustic cavitation \cite{blake,apfel}.  Bubbles forced at pressures
exceeding the Blake threshold grow quasistatically without bound.
This criterion is especially important for gas bubbles in liquids when
surface tension is the dominant effect, such as submicron air bubbles
in water, where the natural oscillation frequencies are high.

In contrast, when the acoustic pressure fields are not quasistatic,
bubbles generally evolve in highly nonlinear fashions
\cite{plesset-prosperetti,feng-leal,leal-book,leighton}.  To begin
with, the intrinsic oscillations of spherically symmetric bubbles in
inviscid incompressible liquids are nonlinear \cite{leal-book}.  The
phase portrait of the Rayleigh-Plesset equation \cite{szeri-leal,
smereka, chang-chen}, consists of a large region of bounded, stable
states centered about the stable equilibrium radius.  The natural
oscillation frequencies of these states depend on the initial bubble
radius and its radial momentum, and this family of states limits on a
state of infinite period, namely a homoclinic orbit in the phase
space, which acts as a boundary outside of which lie initial
conditions corresponding to unstable bubbles.  Time-dependent acoustic
pressure fields then interact nonlinearly with both the periodic
orbits and the homoclinic orbit.  In particular, they can act to break
the homoclinic orbit, permitting initially stable bubbles to leave the
stable region and grow without bound.  These interactions have been
studied from many points of view: experimentally, numerically, and
analytically via perturbation theory and techniques from dynamical
systems.

In \cite{szeri-leal}, the transition between regular and chaotic
oscillations, as well as the onset of rapid radial growth, is studied
for spherical gas bubbles in time-dependent pressure fields.  There,
Melnikov theory is applied to the periodically- and
quasiperiodically-forced Rayleigh-Plesset equation for bubbles
containing an isothermal gas.  One of the principal findings is that,
when the acoustic pressure field is quasiperiodic in time with two or
more frequencies, the transition to chaos and the threshold for rapid
growth occur at lower amplitudes of the acoustic pressure field than
in the case of single-frequency forcing.  Their work was motivated in
turn by that in \cite{kang-leal}, where Melnikov theory was used to
study the time-dependent shape changes of gas bubbles in time-periodic
axisymmetric strain fields.

The work in \cite{smereka}
identifies a rich bifurcation superstructure for radial oscillations
for bubbles in time-periodic acoustic pressure fields.  Techniques
from perturbation theory and dynamical systems are used to analyze
resonant subharmonics, period-doubling bifurcation sequences, the
disappearance of strange attractors, and transient chaos in the
Rayleigh-Plesset equation with small-amplitude liquid viscosity and
isentropic gas.  The analysis in \cite{smereka} complements the
experiments of \cite{esche}
and the experiments and numerical simulations of 
\cite{lauterborn,lauterborn-koch,parlitz}.  
Analyzing subharmonics, these works quantify the impact of increasing
the amplitude of the acoustic pressure field on the frequency-response
curves.

Other works examining the threshold for acoustic cavitation in
time-dependent pressure fields have focused on the case of a step
change in pressure.  In \cite{chang-chen}, the response of a gas
bubble to such a step change in pressure is analyzed by numerical and
Melnikov perturbation techniques to find a correlation between the
cavitation pressure and the viscosity of the liquid.  One of the
principal findings is that the cavitation pressure scales as the
one-fifth power of the liquid viscosity.  A general method to compute
the critical conditions for an instantaneous pressure step is also
given in \cite{dugue}.  The results extend numerical simulations of
\cite{matsumoto} and experimental findings of \cite{sato}, and apply
for any value of the polytropic gas exponent.

The goal of the present article is to apply similar perturbation
methods and techniques from the theory of nonlinear dynamical systems
to refine the Blake cavitation threshold for isothermal bubbles whose
radii are slightly smaller than the critical Blake radius and whose
motions are not quasistatic.  Specifically, we suppose these bubbles
are subjected to time-periodic acoustic pressure fields and,
by reducing the Rayleigh-Plesset equations to a simpler distinguished
limit equation, we obtain the dynamic cavitation threshold for these
subcritical bubbles.

The paper is organized as follows.  In the remainder of this section,
the standard Blake cavitation threshold is briefly reviewed. This also
allows us to identify the critical radius which separates stable and
unstable bubbles that are in equilibrium.
In section \ref{Tdle}, the distinguished limit (or normal
form) equation of motion for subcritical bubbles ({\em i.e.}, those
whose radii is slightly smaller than the critical value) is obtained
from the Rayleigh-Plesset equation.  This necessitates identifying the
natural timescale of oscillation of such subcritical bubbles which
happens to depend upon how close they are to the critical size.  
We begin section \ref{Aft} by defining a simple criterion for determining
when cavitation has occurred.  We then analyze the normal form equation 
and determine the cavitation threshold for a specific value of the acoustic 
forcing frequency (at which the corresponding linear undamped system would
resonate).  This pressure threshold is then compared to numerical
simulations of the full Rayleigh-Plesset equation and the good
agreement found between the two is demonstrated.  The self-consistency
of the distinguished limit equation is further discussed in that
section.  
Section \ref{Ptfgo} generalizes the results to include arbitrary
acoustic forcing frequencies.  Acoustic forcing frequencies which
facilitate cavitation using the least forcing pressure are determined.
An unusual dependence of the threshold pressure on forcing frequency is
discovered and explained by analyzing the ``slowly-varying'' phase-plane 
of the dynamical system.  
At the end of section \ref{Ptfgo}, our choice of a cavitation criterion is
discussed in the setting of a Melnikov analysis.
In section \ref{Pfwtff} we extend the cavitation results to the case of an 
oscillating subcritical bubble that is driven simultaneously at two 
different frequencies.
We recap the paper in section \ref{Discussion} by highlighting the main 
results and discussing their applicability.  Lastly, we conclude the paper 
with an appendix which qualitatively discusses the relation of our results
to some recent experimental findings.

\subsection{Blake threshold pressure}
\label{Btp}

To facilitate the development of subsequent sections we first briefly
review the derivation of the Blake threshold \cite{leighton}.  At
equilibrium, the pressure, $p_{B}$, inside a spherical bubble of
radius $R$ is related to the pressure, $p_{L}$, of the outside liquid
through the normal stress balance across the surface:
\begin{equation}
p_{B}=p_{L}+\frac{2 \sigma}{R}.
\label{stress_balance}
\end{equation}
The pressure inside the bubble consists of gas pressure and vapor
pressure, $p_{B}=p_{g}+p_{v}$, where the vapor pressure $p_v$ is taken
to be constant --- $p_v$ depends primarily on the temperature of the
liquid --- and the pressure of the gas is assumed to be given by the
equation of state:
\begin{equation}
p_{g}=p_{g_{0}} \left( \frac{R_{0}}{R} \right)^{3 \gamma},
\label{gas_pressure1}
\end{equation}
with $\gamma$ the polytropic index of the gas. For isothermal
conditions $\gamma=1$, whereas for adiabatic ones, $\gamma$ is the
ratio of constant-pressure to constant-volume heat capacities.  At
equilibrium, the bubble has radius $R_{0}$, the gas has pressure
$p_{g_{0}}$ and the static pressure of the liquid is taken to be
$p_{0}^{\infty}$.  Thus, the equilibrium pressure of the gas in the
bubble is given by
\begin{displaymath}
p_{g_{0}} = p_{0}^{\infty} - p_{v} + \frac{2 \sigma}{R_{0}}.
\end{displaymath}
Upon substituting this result into (\ref{gas_pressure1}) we get the
following expression for the pressure of the gas inside the bubble as
a function of the bubble radius:
\begin{equation}
p_{g}=\left( p_{0}^{\infty} - p_{v} + \frac{2 \sigma}{R_{0}} \right)
\left( \frac{R_{0}}{R} \right)^{3 \gamma}.
\label{gas_pressure2}
\end{equation}
Upon combining equations (\ref{stress_balance}) and
(\ref{gas_pressure2}), we find
\begin{equation}
p_{L} =
\left( p_{0}^{\infty} - p_{v} + \frac{2 \sigma}{R_{0}} \right)
\left( \frac{R_{0}}{R} \right)^{3 \gamma}
+ p_{v} - \frac{2 \sigma}{R}.
\label{liquid_pressure}
\end{equation}
Equation (\ref{liquid_pressure}) governs the change in the radius of a
bubble in response to quasistatic changes in the liquid pressure
$p_L$.  More precisely, by ``quasistatic'' we mean that the liquid
pressure changes slowly and uniformly with inertial and viscous
effects remaining negligible during expansion or contraction of the
bubble.  For very small (sub-micron) bubbles, surface tension is the
dominant effect.  Furthermore, typical acoustic forcing frequencies
are much smaller than the resonance frequencies of such tiny bubbles.
In this case, the pressure in the liquid changes very slowly and
uniformly compared to the natural timescale of the bubble.

\begin{figure}[h!]
\begin{center}
\epsfig{figure=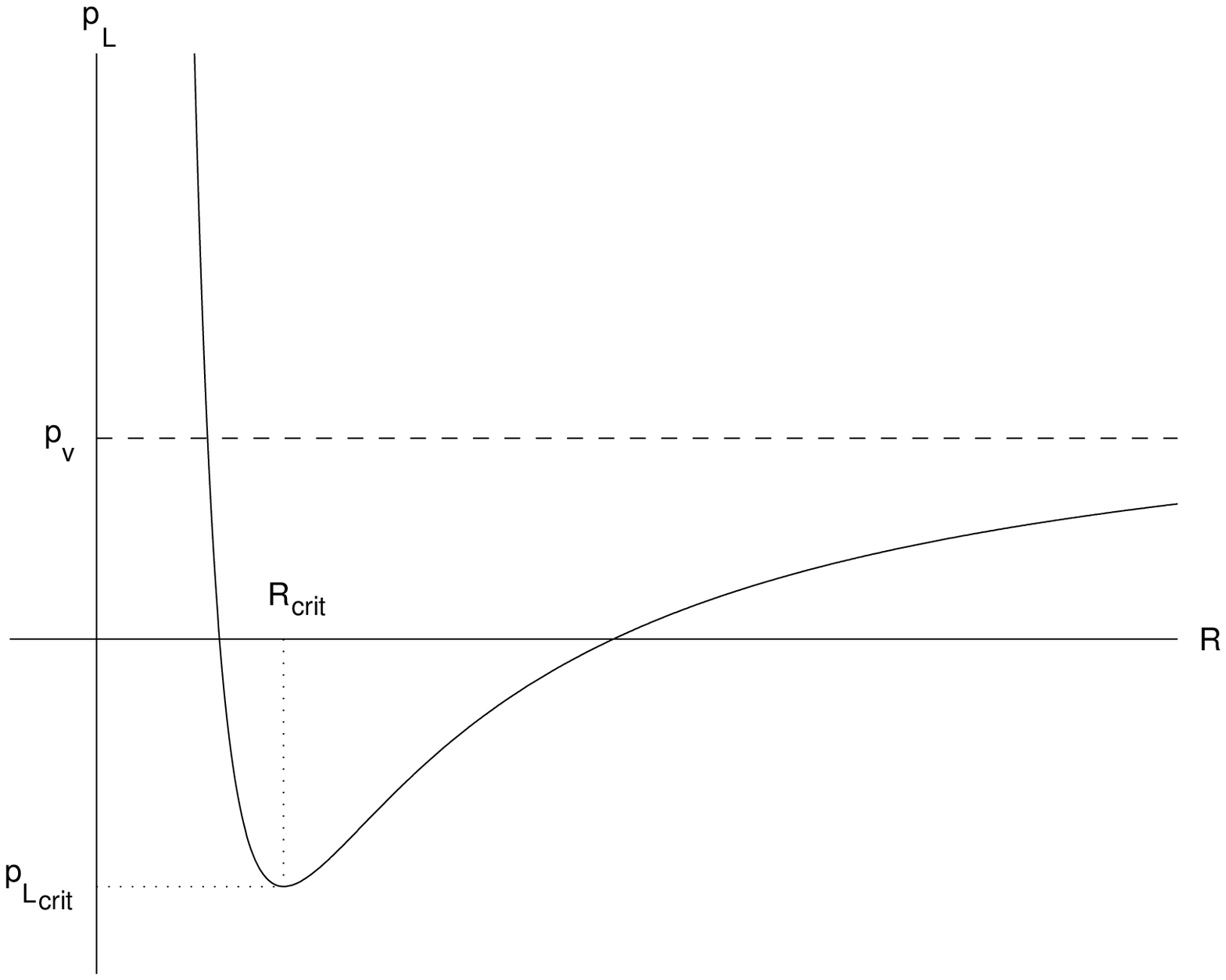,height=5in,width=5in}
\end{center}
\caption{}
\label{figure1}
\end{figure}

For very small bubbles, the Peclet number for heat transfer within the
bubble --- defined as $R_0^2 \omega / \alpha$, with $\omega$ the
bubble natural frequency (see subsection \ref{Derivation}) and $\alpha$ 
the thermal diffusivity of the gas --- is small, and due to the rapidity 
of thermal conduction over such small length scales, the bubble may be 
regarded as isothermal.  We therefore let $\gamma = 1$ for an isothermal 
bubble and define
\begin{displaymath}
\tilde{G} =
\left( p_{0}^{\infty} - p_{v} + \frac{2 \sigma}{R_{0}} \right)
R_{0}^{3}.
\end{displaymath}
Then equation (\ref{liquid_pressure}) becomes
\begin{equation}
p_{L}=p_{v}+\frac{\tilde{G}}{R^{3}}- \frac{2 \sigma}{R}.
\label{liquid_pressure_iso}
\end{equation}
The right-hand side of this equation is plotted in figure
\ref{figure1} (solid curve), which shows a minimum value at a critical
radius labeled $R_{\rm crit}$.

Obviously, if the liquid pressure is lowered to a value below the
corresponding critical pressure $p_{L_{\rm crit}}$, no equilibrium
radius exists. For values of $p_L$ which are above the critical value
but below the vapor pressure $p_v$, equation
(\ref{liquid_pressure_iso}) yields two possible solutions for the
radius $R$. Bubbles whose radii are less than the Blake radius,
$R_{\rm crit}$, are stable to small disturbances, whereas bubbles with
$R > R_{\rm crit}$ are unstable to small disturbances.  

The Blake radius itself can be obtained by finding the minimum
of the right-hand side of (\ref{liquid_pressure_iso}) for $R > 0$.
This yields the critical Blake radius
\begin{equation}
R_{\rm crit} = \left( \frac{3 \tilde{G}}{2 \sigma} \right)^{1/2},
\label{R_crit}
\end{equation}
at which the corresponding critical liquid pressure is
\begin{equation}
p_{L_{\rm crit}} = p_{v} -
\left( \frac{32 \sigma^{3}}{27 \tilde{G}} \right)^{1/2}.
\label{p_L_crit}
\end{equation}
By combining the last two equations, it is also possible to express
the Blake radius in the form:
\begin{displaymath}
R_{\rm crit} = \frac{ 4 \sigma }{3 (p_v - p_{L_{\rm crit}})} \,,
\end{displaymath}
relating the critical bubble radius to the critical pressure in the
liquid.  Bubbles whose radii are smaller than $R_{\rm crit}$ are
quasistatically stable, while bigger ones are unstable.

To obtain the standard Blake pressure we assume that $p_{v}$ can be
ignored and recall that surface tension dominates in the quasistatic
regime which amounts to $p_{0}^{\infty} \ll 2 \sigma / R_{0}$.  Under
these approximations, $\tilde{G}\approx 2 \sigma R_0^2$ and the Blake
threshold pressure is conventionally defined as
\begin{eqnarray*}
p_{\rm Blake} & \equiv & p_{0}^{\infty} - p_{L_{\rm crit}} \\
              & \approx & p_{0}^{\infty} + 0.77 \frac{\sigma}{R_{0}}.
\end{eqnarray*}
In the quasistatic regime where the Blake threshold is valid, $p_{\rm
Blake}$ is the amplitude of the low-frequency acoustic pressure beyond
which acoustic forcing at higher pressures is sure to cause
cavitation.  When the pressure changes felt by the bubble are no
longer quasistatic, a more detailed analysis taking into
consideration the bubble dynamics and acoustic forcing frequency must
be performed to determine the cavitation threshold. This is the type
of analysis we undertake in this contribution.

\section{The distinguished limit equation}
\label{Tdle}

\subsection{Derivation}
\label{Derivation}

To make progress analytically, we focus our attention on
``subcritical'' bubbles whose radii are only slightly smaller than the
Blake radius at a given liquid pressure below the vapor pressure. We
thus define a small parameter $\epsilon > 0$ by
\begin{equation}
\epsilon =  2 \left[1- \frac{R_{0}}{R_{\rm crit}} \right] ,
\label{eps}
\end{equation}
which measures how close the equilibrium bubble radius $R_0$ is to the
critical value $R_{\rm crit}$.  The value of the mean pressure in the
liquid, corresponding to the equilibrium radius $R_0$, can also be
found from equation (\ref{liquid_pressure_iso}) to be
\begin{equation}
p_0^\infty - p_v = \frac{2 \sigma}{3 R_0}\,[(1-\frac{\epsilon}{2})^{-2}
- 3] = -\frac{4 \sigma}{3 R_0}\,[1-\frac{1}{2} \epsilon - 
\frac{3}{8} \epsilon^{2} + {\cal O}(\epsilon^{3})] \,.
\label{subcritliqpress}
\end{equation}
The liquid pressure $p_0^\infty$ and the critical pressure $p_{L_{\rm
crit}}$ differ only by an ${\cal O}(\epsilon^{2})$ amount.

It turns out that the characteristic time scale for the natural
response of such subcritical bubbles also depends on the small
parameter $\epsilon$.  This timescale for small amplitude oscillations
of a spherical bubble is obtained by linearizing the isothermal,
unforced Rayleigh-Plesset equation \cite{plesset-prosperetti}
\begin{equation}
\rho \left[ R \ddot{R} + \frac{3}{2} \dot{R}^{2} \right]
=
\left( p_{0}^{\infty} - p_{v} + \frac{2 \sigma}{R_{0}} \right)
\left( \frac{R_{0}}{R} \right)^{3}
+ p_{v} - \frac{2 \sigma}{R} - p_{0}^{\infty},
\label{RP_sect3}
\end{equation}
where the density of the liquid is given by $\rho$ and viscosity has
been neglected.  Specifically, we substitute $R=R_{0}(1+x)$ into
(\ref{RP_sect3}) and keep terms linear in $x$ to get:
\begin{equation}
\ddot{x} + \left[ 
\frac{4 \sigma}{\rho R_{0}^{3}} +
\frac{3 (p_{0}^{\infty}-p_{v})}{\rho R_{0}^{2}}
\right] x = 0.
\label{linear_RP_sect3}
\end{equation}
Solutions to (\ref{linear_RP_sect3}), representing small amplitude
oscillations about equilibrium, are therefore $x=x_0\cos(\omega
t+\phi)$ with the angular frequency given by
\begin{equation}
\omega = \left[ 
\frac{4 \sigma}{\rho R_{0}^{3}} +
\frac{3 (p_{0}^{\infty}-p_{v})}{\rho R_{0}^{2}}
\right]^{1/2}.
\label{natural_freq}
\end{equation}
We now use $\omega$ to define a nondimensional time variable: $\tau =
\omega t$.  We are interested in analyzing stability for values of
$(R_{0},p_{0}^{\infty})$ near $(R_{\rm crit},p_{L_{\rm crit}})$.
Hence, upon recalling (\ref{R_crit}), (\ref{p_L_crit}) and (\ref{eps}),
we see that:
\begin{displaymath}
\tau = \left[
\frac{2 \sigma}{\rho R_{0}^{3}}
\left( 2 \left[1- \frac{R_{0}}{R_{\rm crit}} \right] \right)
\right]^{1/2} t =
\left[
\frac{2 \sigma \epsilon}{\rho R_{0}^{3}}
\right]^{1/2} t \,.
\end{displaymath}
We note that as $\epsilon$ tends to zero, the timescale for bubble
oscillations (the reciprocal of the factor multiplying $t$ in the last
equation) increases as $\epsilon^{-1/2}$.

Having determined the proper scaling for the time variable for
slightly subcritical bubbles, we can now find the distinguished limit
(or normal form) equation for such bubbles in a time-periodic pressure
field.  We start with the isothermal, viscous Rayleigh-Plesset
equation \cite{plesset-prosperetti}:
\begin{equation}
\rho \left[ R \ddot{R} + \frac{3}{2} \dot{R}^{2} \right]
+ 4 \mu \frac{\dot{R}}{R}
=
\left( p_{0}^{\infty} - p_{v} + \frac{2 \sigma}{R_{0}} \right)
\left( \frac{R_{0}}{R} \right)^{3}
+ p_{v} - \frac{2 \sigma}{R} - p_{0}^{\infty} + p_{A} \sin({\Omega t}).
\label{rp_eqn}
\end{equation}
The amplitude and frequency of the applied acoustic forcing are given
by $p_{A}$ and $\Omega$, respectively, and $\mu$ represents the
viscosity of the fluid. Here, the far-field pressure in the liquid has
been taken to be $p_0^\infty - p_A \sin(\Omega t)$, with $p_0^\infty$
given by equation (\ref{subcritliqpress}).  Setting
$R(t)=R_{0}(1+\epsilon x(\tau))$, with $\epsilon$ the same small
parameter introduced above, we obtain at order $\epsilon^2$ (noting
that all of the ${\cal O}(1)$ and ${\cal O}(\epsilon)$ terms cancel):
\begin{equation}
\ddot{x} + 2 \zeta \dot{x} + x - x^{2} = A \sin(\Omega^{*} \tau),
\label{dist_lim_eqn}
\end{equation}
where
\begin{equation}
\zeta = \left( \frac{2 \mu^{2}}{\epsilon \sigma \rho R_{0}} \right)^{1/2},
\;\;\;
A = \frac{p_{A} R_{0}}{2 \sigma \epsilon^{2}},
\;\;\;
\Omega^{*} = \Omega 
\left( \frac{\rho R_{0}^{3}}{2 \sigma \epsilon} \right)^{1/2}.
\label{dist_lim_params}
\end{equation}
In equation (\ref{dist_lim_eqn}), each overdot represents a derivative
with respect to $\tau$.  

It is implicit in the above scaling that
$\zeta$, $A$, and $\Omega^{*}$ are nondimensional and ${\cal O}(1)$
with respect to $\epsilon$.  To see that this is reasonable, consider
an air bubble in water with 
$\rho=998$ kg/m$^{3}$,
$\mu=0.001$ kg/m$\cdot$s,
$\sigma=0.0725$ N/m.
If we specify $\epsilon = 0.1$ and take a modest equilibrium radius
of $R_{0} = 2 \times 10^{-6} \,$m then $\zeta = 0.38$.  Our analysis
of (\ref{dist_lim_eqn}) in subsequent sections will concentrate
primarily on values of $\zeta$ in the range $0 \le \zeta \le 0.4$.
The parameters $A$ and $\Omega^{*}$ are related to the forcing conditions,
and their magnitudes can be made order unity by choosing appropriate
forcing parameters $p_A$ and $\Omega$.  As an example, if we again
choose $R_{0}$ to equal 2 microns, then
$\Omega^{*}=(2.35 \times 10^{-7}$s$)\Omega/\sqrt{\epsilon}$.
Moreover, setting $\epsilon=0.05$ gives
$\Omega^{*}=(1.05 \times 10^{-6}$s$)\Omega$.  Hence, the dimensionless
parameter $\Omega^{*}$ is ${\cal O}(1)$ when $\Omega$ is in the
megahertz range, and this is precisely the frequency range we are
interested in exploring.  Similarly, with $R_{0}$ and $\sigma$ chosen
as above, we find 
$A=(1.38 \times 10^{-5}$m$\cdot$s$^{2}/$kg$)p_{A}/\epsilon^{2}$
and thereby we see that if $\epsilon=0.1$ then $p_{A}$ can become on the
order of $10^{3}$Pa.  More data will be presented later, in figure 
\ref{figure9}, showing typical forcing pressures.

\subsection{Interpretation}
\label{Interpretation}

In the laboratory one can create a subcritical bubble by subjecting
the liquid to a low-frequency transducer whose effect is to lower
the ambient pressure below the vapor pressure.  Then a second
transducer of high frequency (high relative to the slow transducer)
will give rise to the forcing term on the right hand side of
(\ref{dist_lim_eqn}).
The low-frequency transducer periodically increases and decreases the
pressure in the liquid (and shrinks and expands the bubble, which
follows this pressure field quasistatically). When the peak negative
pressure is reached (and the bubble has expanded to its maximum size),
we can imagine that state as the new equilibrium state, and at that
point bring in the effects of the sound from the second transducer.
This second field can then possibly make the bubble, which had already
grown to some large size (but still smaller than the critical radius),
become unstable.  This would all happen very fast compared to the time
scale of the original slow transducer, so the pressure field
contributed by the original transducer remains near its most negative
value throughout.  The stability response of the bubble to the high
frequency component of the pressure field is the subject of the rest
of this work.

\section{Acoustic forcing thresholds ($\Omega^{*}=1$)}
\label{Aft}

The value of $\Omega^*=1$ corresponds to the forcing frequency at
which the linear and undamped counterpart of (\ref{dist_lim_eqn})
would resonate. We therefore choose this value of the forcing
frequency as a starting point and perform a detailed analysis of
the dynamics inherent in the distinguished limit equation at
this value of $\Omega^*$.
We caution, however, that, as with most forced, damped nonlinear
oscillators, the largest resonant response occurs away from the
resonance frequency of the linear oscillator.  We use $\Omega^{*}=1$
mainly as a starting point for the analysis, and the dynamics observed
for a range of other $\Omega^{*}$ values is reported in section \ref{Ptfgo}.

\begin{figure}[h!]
\begin{center}
\begin{minipage}{2.7in}
\epsfig{figure=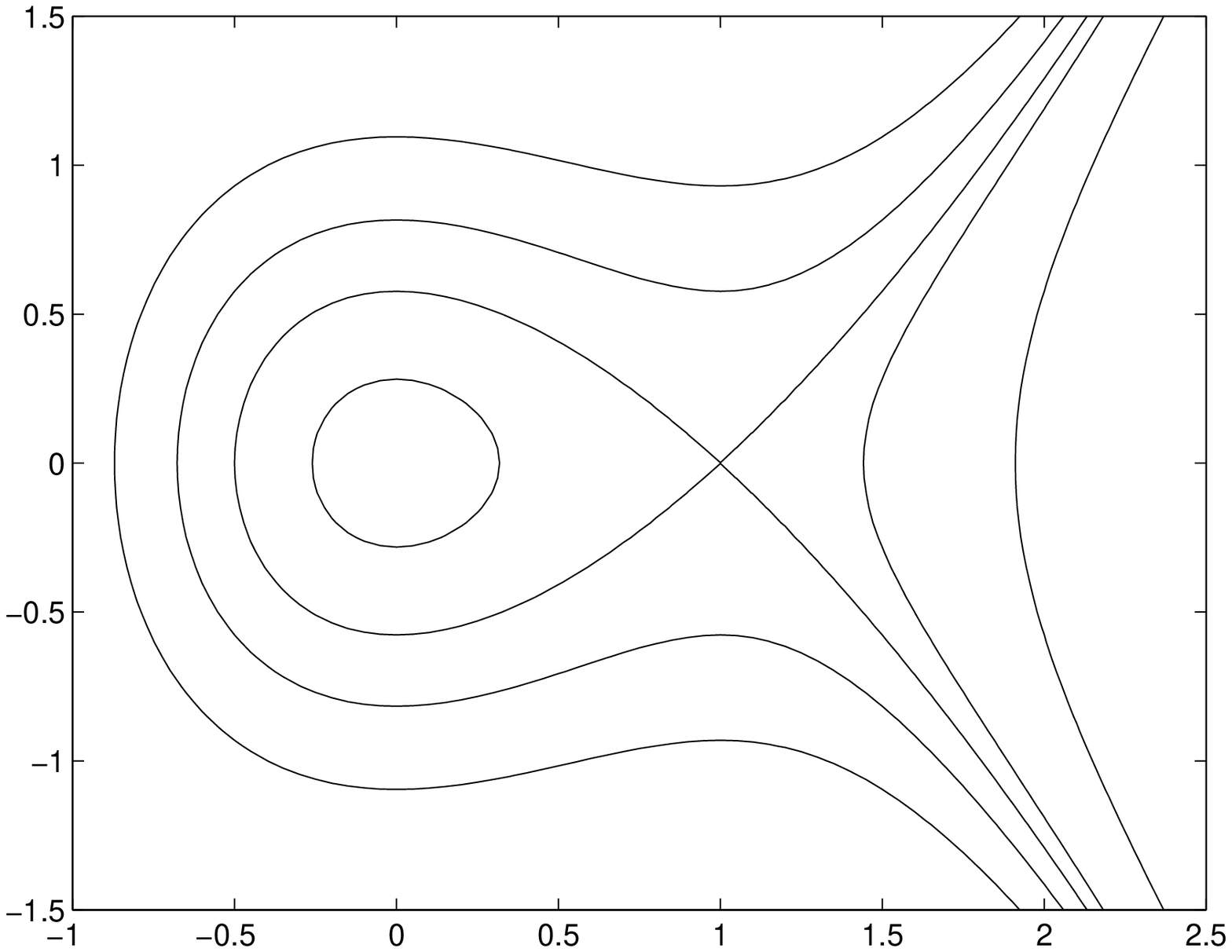,height=2.7in,width=2.7in}
\begin{center} (a) \end{center}
\end{minipage}
\hspace*{0.1in}
\begin{minipage}{2.7in}
\epsfig{figure=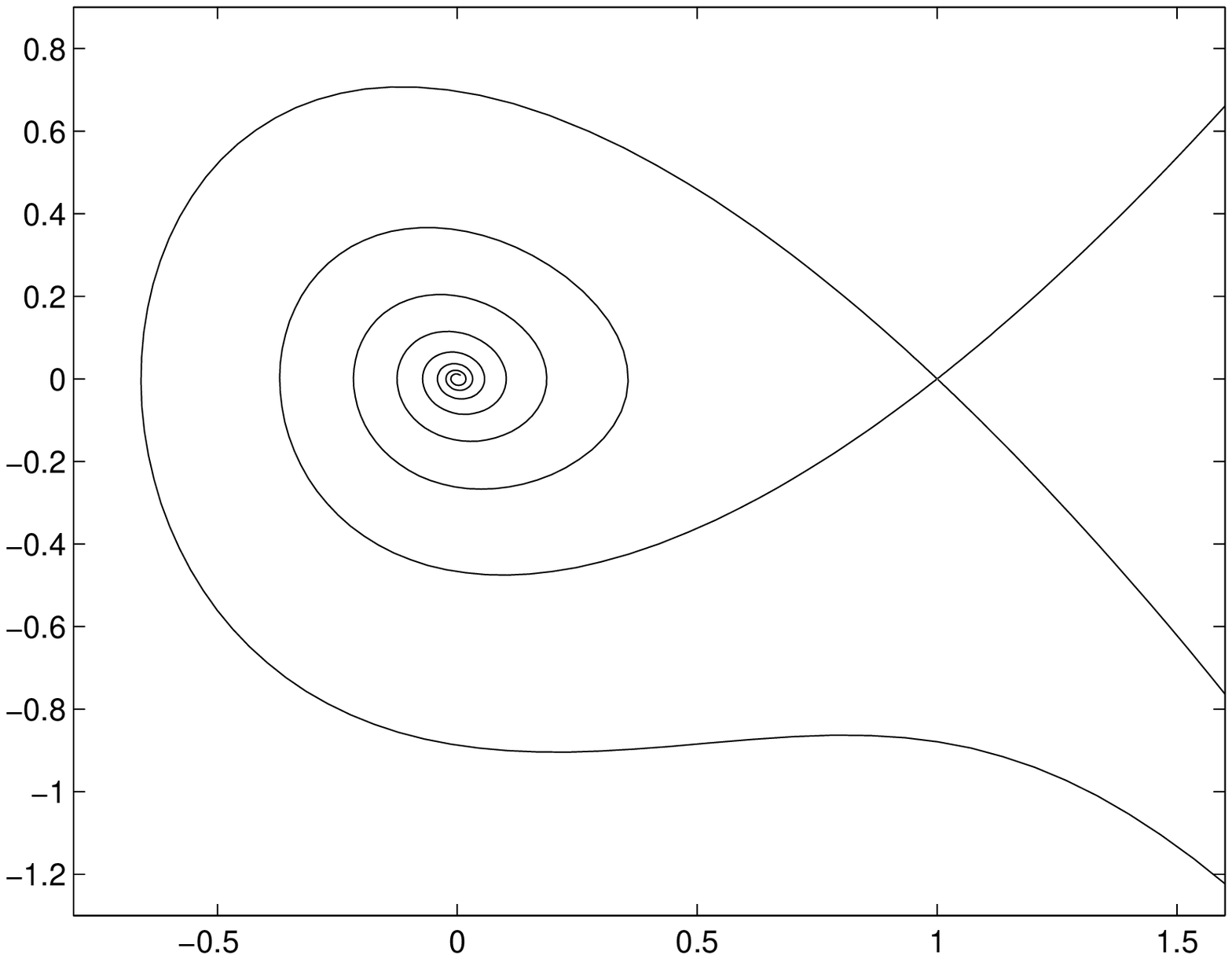,height=2.7in,width=2.7in}
\begin{center} (b) \end{center}
\end{minipage}
\end{center}
\caption{}
\label{figure2}
\end{figure}

\begin{figure}[h!]
\begin{center}
\epsfig{figure=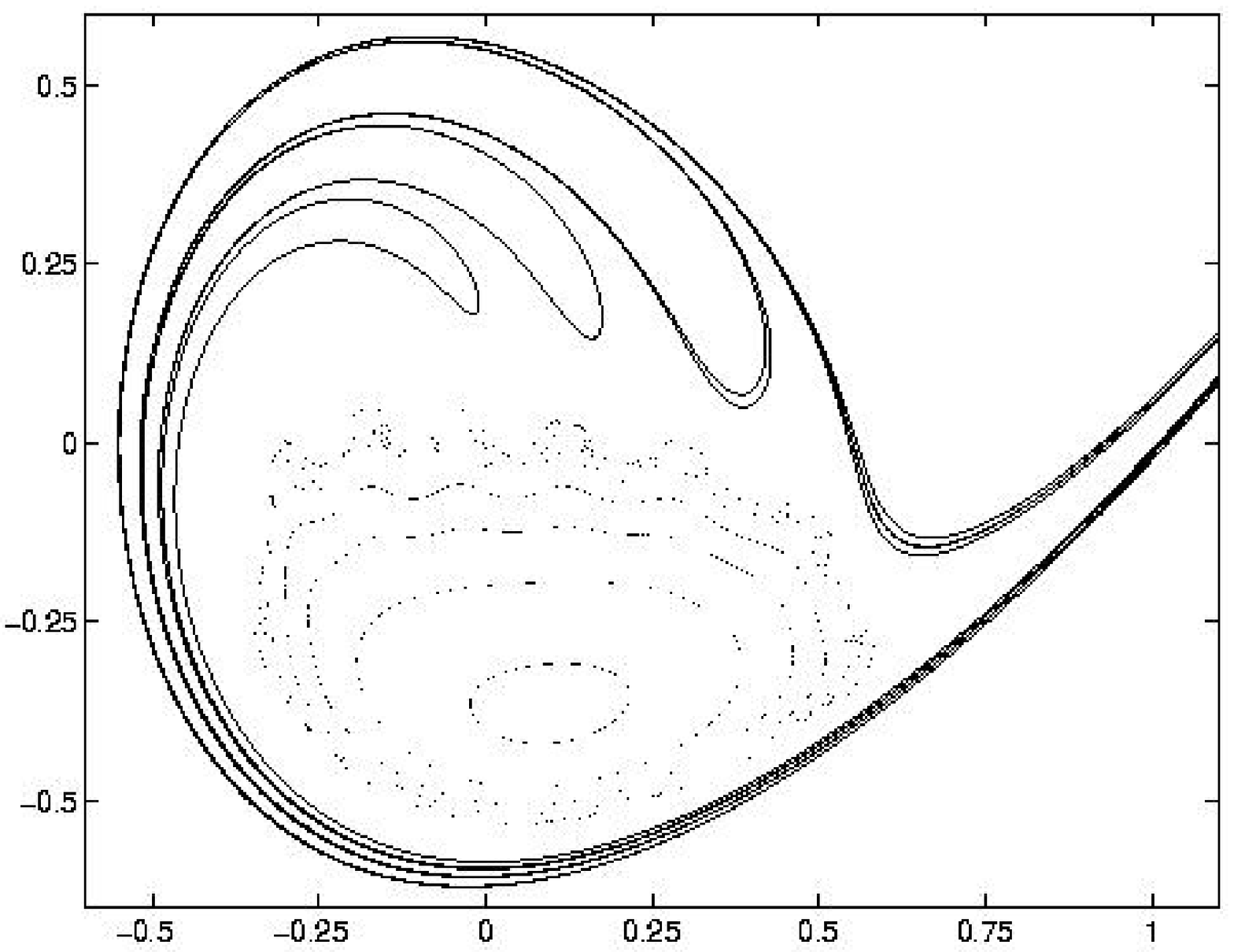,height=5in,width=5in}
\end{center}
\caption{}
\label{figure3}
\end{figure}

Some special cases of (\ref{dist_lim_eqn}) can be readily analyzed
when $\Omega^{*}=1$. In the absence of forcing, {\em i.e.}, when
$A=0$, the phase portraits of (\ref{dist_lim_eqn}) with $\zeta \geq 0$
are shown in figure \ref{figure2}. With no damping 
(figure \ref{figure2}a), the phase plane has a saddle point at (1,0) and
a center at (0,0).  The latter represents the equilibrium radius of
the bubble which, when infinitesimally perturbed, results in simple
harmonic oscillations of the bubble about that equilibrium. The saddle
point at (1,0) represents the effects of the second nearby root of the
equation (\ref{liquid_pressure_iso}) which is an unstable equilibrium
radius.  When damping is added (figure
\ref{figure2}b), the saddle point remains a saddle, but the center at
(0,0) becomes a stable spiral, attracting a well-defined region of the
phase space towards itself. In the presence of weak forcing (small
$A$) but with no damping ($\zeta=0$), the behavior of
(\ref{dist_lim_eqn}) can be seen in a Poincar\'{e} section shown in
figure \ref{figure3}.

\subsection{Phase plane criterion for acoustic cavitation}
\label{Ppcfac}

\begin{figure}[h!]
\begin{center}
\epsfig{figure=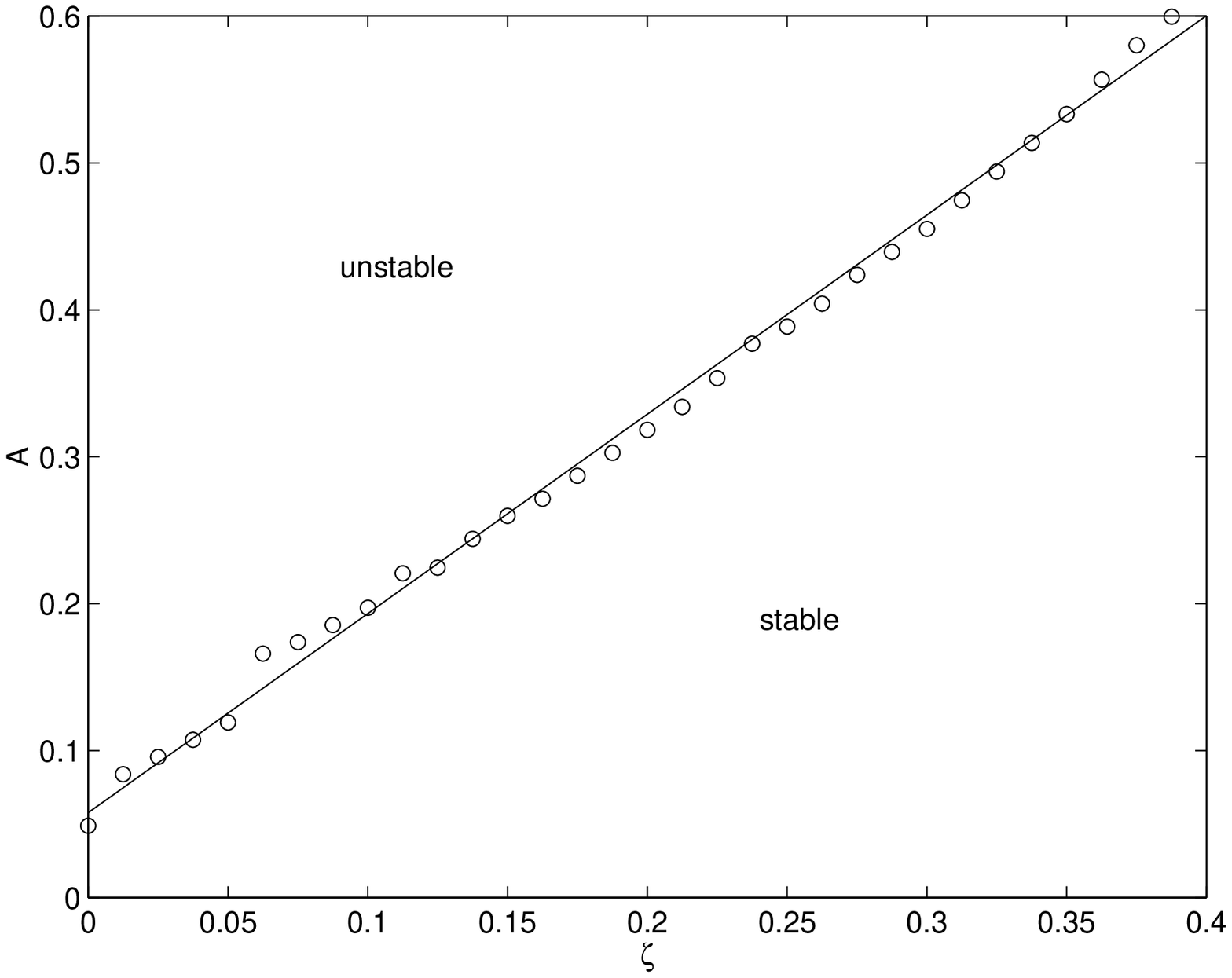,height=5in,width=5in}
\end{center}
\caption{}
\label{figure4}
\end{figure}

To determine when a slightly subcritical bubble becomes unstable we
choose a simple criterion based upon the phase portrait of the
distinguished limit equation (\ref{dist_lim_eqn}).  For a given
$\zeta$, there exists a threshold value, $A_{\rm esc}$, of $A$ such
that the trajectory through the origin (0,0) grows without bound for
$A > A_{\rm esc}$, whereas that trajectory stays bounded for $A <
A_{\rm esc}$.  A stable subcritical bubble becomes unstable as $A$
increases past $A_{\rm esc}$.  Thus there is a stability curve in the
$(A,\zeta)$-plane separating the regions of this parameter space for
which the trajectory starting at the origin in the phase-plane either
escapes to infinity or remains bounded.  Numerically, many such
threshold $\zeta,A_{\rm esc}$ pairs (represented by the open circles
in figure \ref{figure4}) were found with $\Omega^{*}=1$.  The data are
seen empirically to be well fitted by a least-squares straight line,
given by $A_{\rm esc} = 1.356 \zeta + 0.058$.

For practical experimental purposes a linear regression curve based
upon our escape criterion should provide a useful cavitation threshold
for the acoustic pressure in the following dimensional form:
\begin{equation}
p_{A} >
3.835 \frac{\epsilon^{3/2} \sigma^{1/2} \mu}{\rho^{1/2} R_{0}^{3/2}}
+
0.116 \frac{\epsilon^{2} \sigma}{R_{0}}.
\label{stab_thresh}
\end{equation}
Here, $\epsilon$ is given by equation (\ref{eps}) and is itself a
function of the equilibrium radius $R_0$, surface tension $\sigma$ and
the pressure differential $p_0^\infty - p_v$.

\subsection{Period doubling in the distinguished limit equation}
\label{Pditdle}

\begin{figure}[h!]
\begin{center}
\epsfig{figure=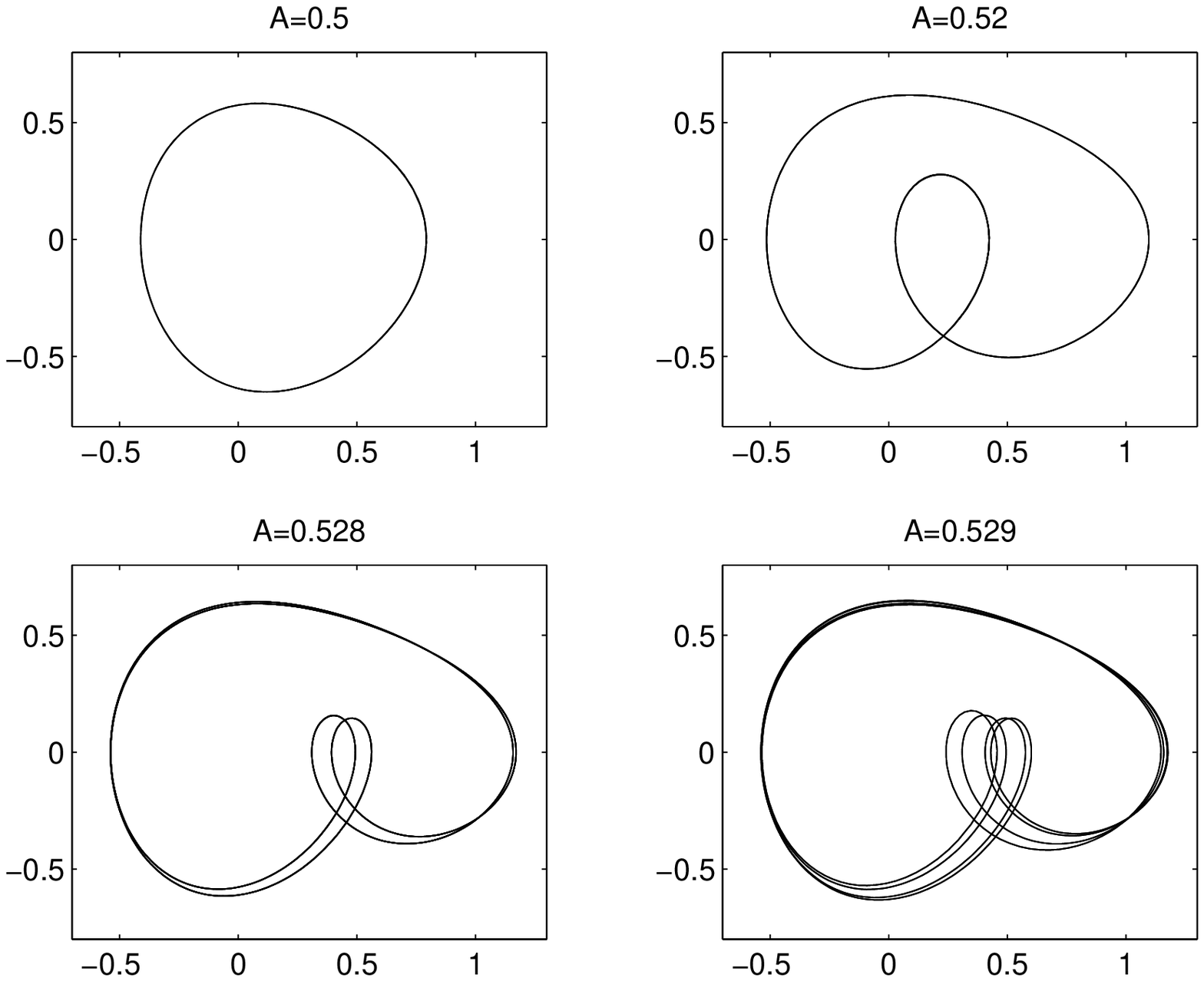,height=5in,width=5in}
\end{center}
\caption{}
\label{figure5}
\end{figure}

\begin{figure}[h!]
\begin{center}
\epsfig{figure=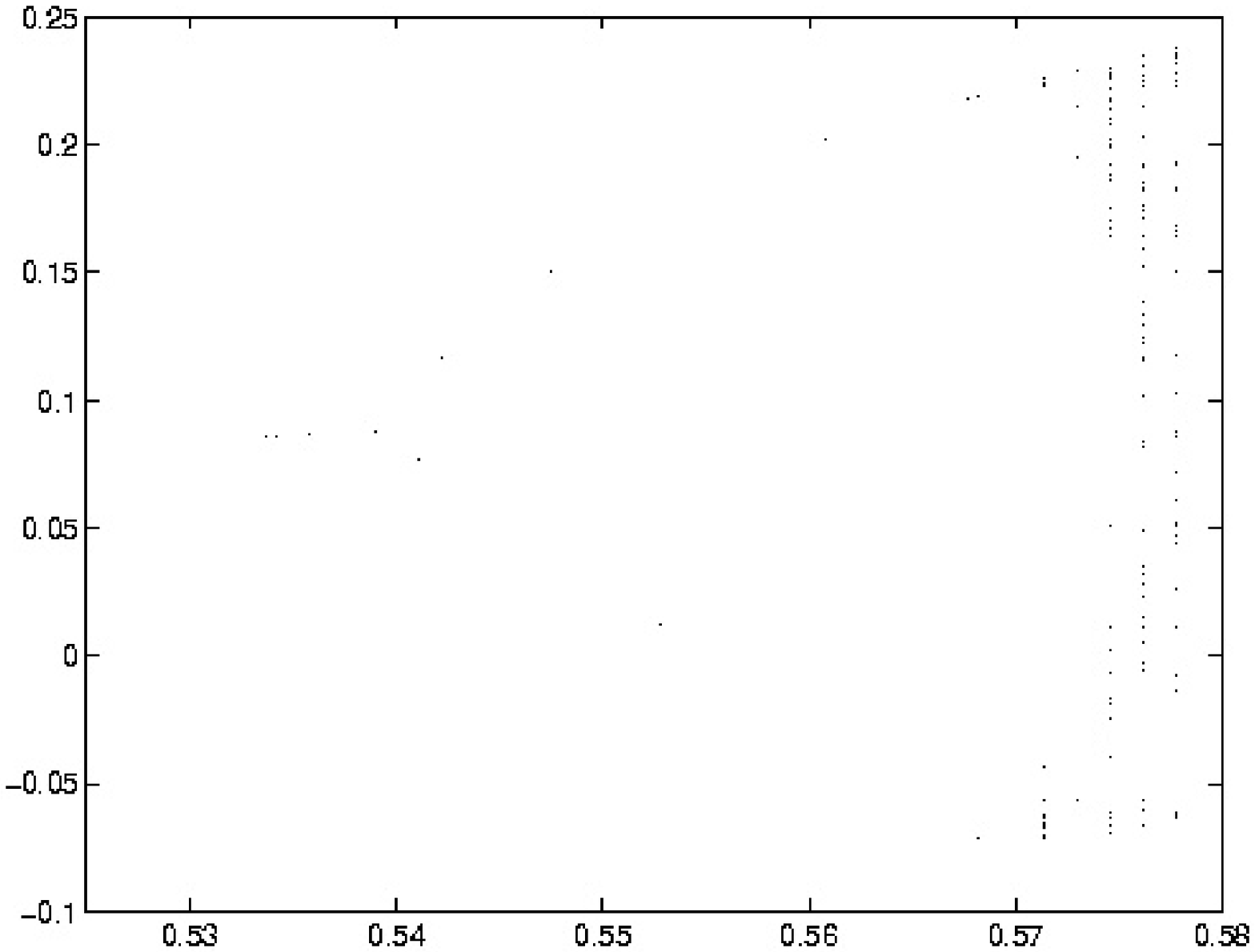,height=5in,width=5in}
\end{center}
\caption{}
\label{figure6}
\end{figure}

It so happens that the stability curve for the trajectory of the
origin can also be interpreted in terms of the period doubling route
to chaos for the escape oscillator (\ref{dist_lim_eqn}).  In other
words, the value of $A_{\rm esc}$ happens to be very near the limiting
value at which the oscillations become chaotic, just before getting
unbounded.  For a fixed value of $\zeta > 0$ and a small enough $A$,
the trajectory of the origin will settle upon a stable limit cycle in
the phase plane.  As $A$ is increased gradually, the period of this
stable limit cycle undergoes a doubling cascade as shown in figure
\ref{figure5} for a fixed value of $\zeta=0.35$.  The period doubling 
sequence will continue as $A$ is increased until the trajectory of the
origin eventually becomes chaotic, but still remains bounded.
Finally, at a threshold value of $A$ the trajectory of the origin will
escape to infinity.  This is the value of $A$ that is given by the
open circles on the stability diagram (figure \ref{figure4}).  A
typical bifurcation diagram for the escape oscillator
(\ref{dist_lim_eqn}) with $\zeta > 0$ is shown in figure
\ref{figure6} in which $\zeta=0.375$. 

\subsection{Robustness of the simple cavitation criterion}
\label{Rotscc}

\begin{figure}[h!]
\begin{center}
\epsfig{figure=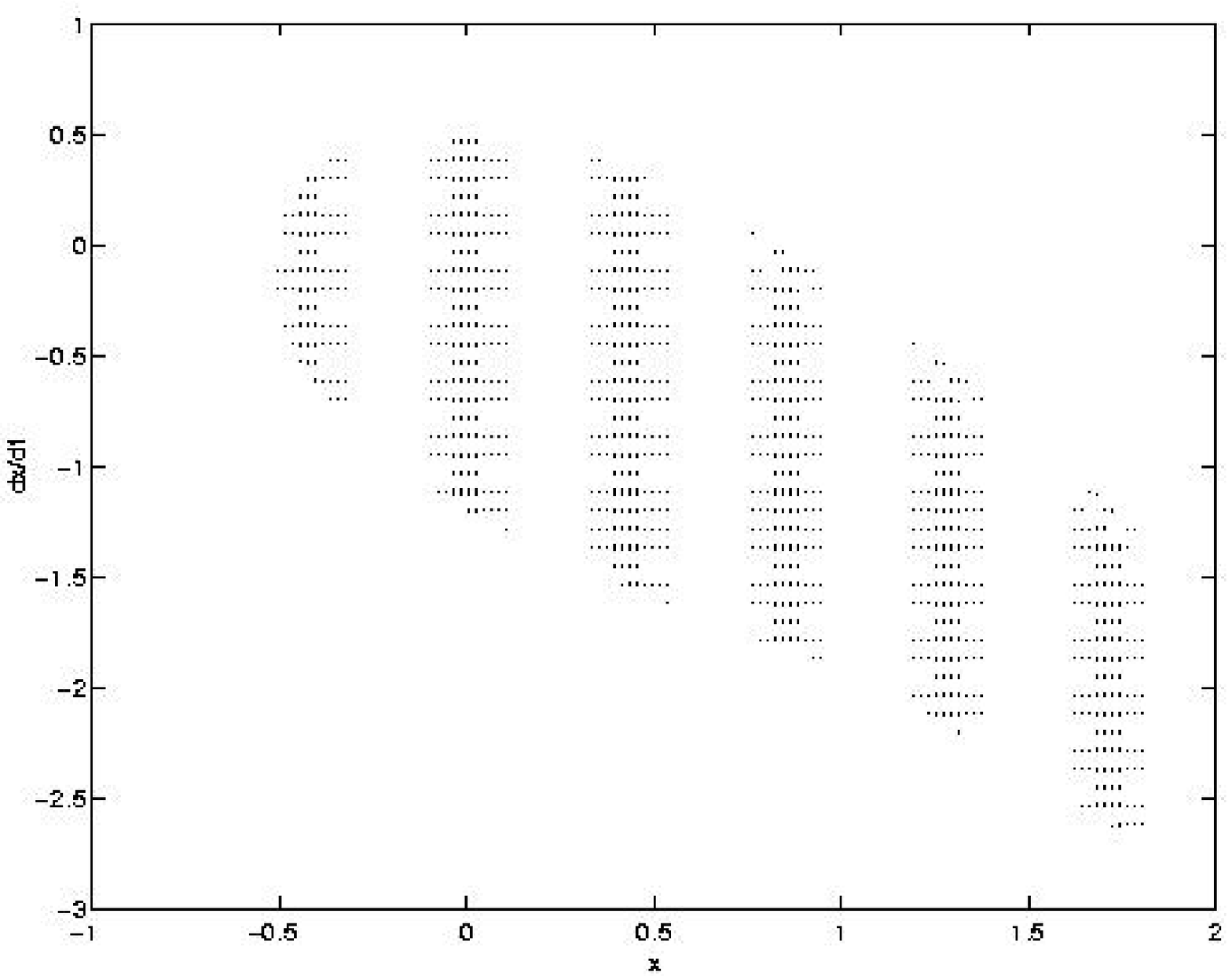,height=5in,width=5in}
\end{center}
\caption{}
\label{figure7}
\end{figure}

In this subsection we justify defining a cavitation criterion
based upon the fate of a single initial condition.
In all simulations with $\zeta>0$, there is a large region of initial
conditions whose fate (escaping or staying bounded) is the same as that 
of the origin (figure \ref{figure7}).  In fact, when the trajectory through 
the origin stays bounded it is clear from the simulations that the origin 
lies in the basin of attraction of a bounded, attracting periodic orbit, 
and points in a large region around it all lie in the basin of attraction 
of the same orbit.  The trajectories through all points in that basin 
remain bounded.  Then, as the forcing amplitude is increased the attractor 
is observed to undergo a sequence of period-doubling bifurcations, and 
this sequence culminates at the forcing magnitude when the origin and 
other initial conditions in a large region about it escape, because there 
is no longer a bounded attracting orbit in whose basin of attraction they lie.

\subsection{Comparison with full Rayleigh-Plesset simulations}
\label{Cwfrps}

\begin{figure}[h!]
\begin{center}
\begin{minipage}{2.25in}
\epsfig{figure=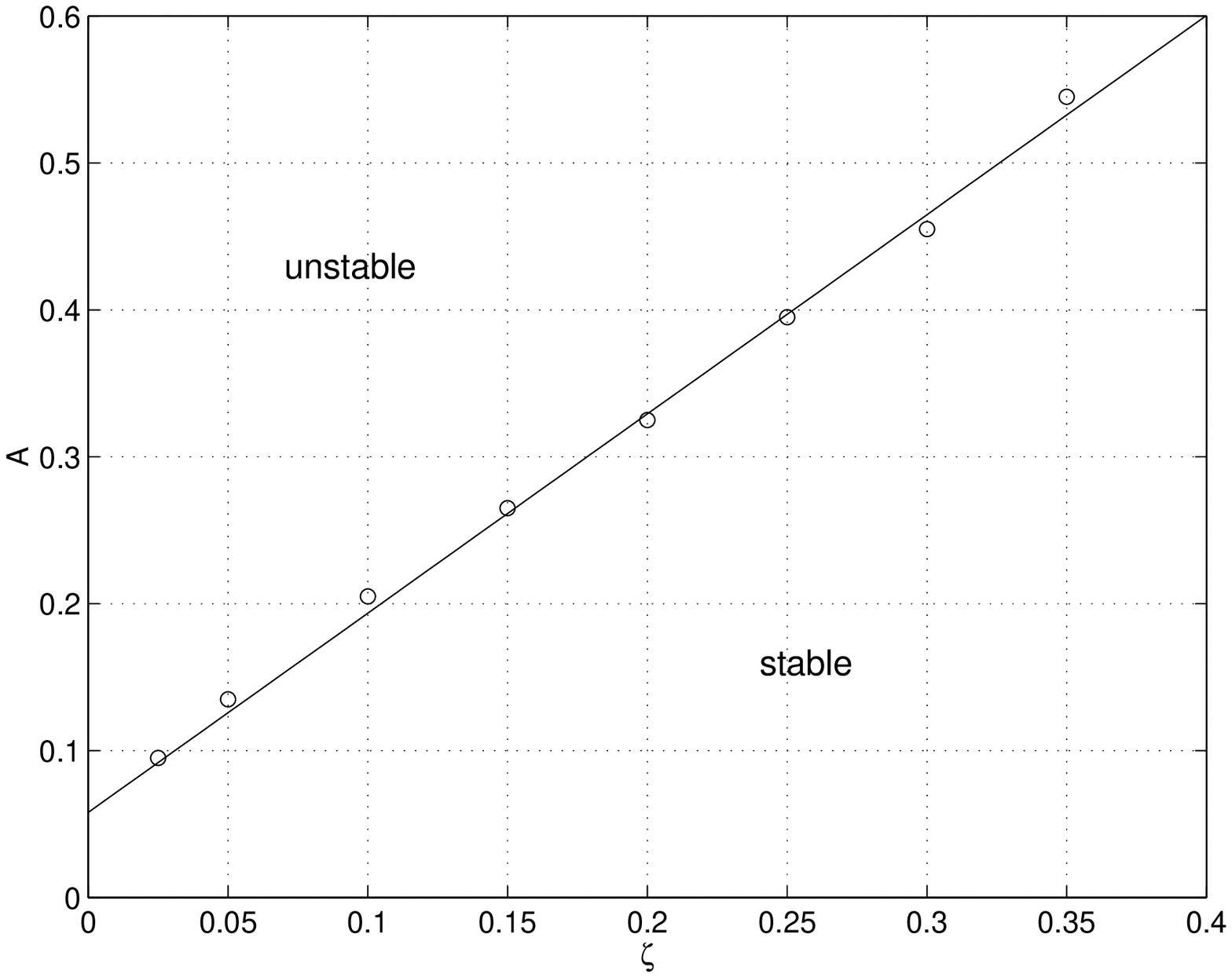,height=2.25in,width=2.25in}
\begin{center} (a) \end{center}
\end{minipage}
\hspace*{0.2in}
\begin{minipage}{2.25in}
\epsfig{figure=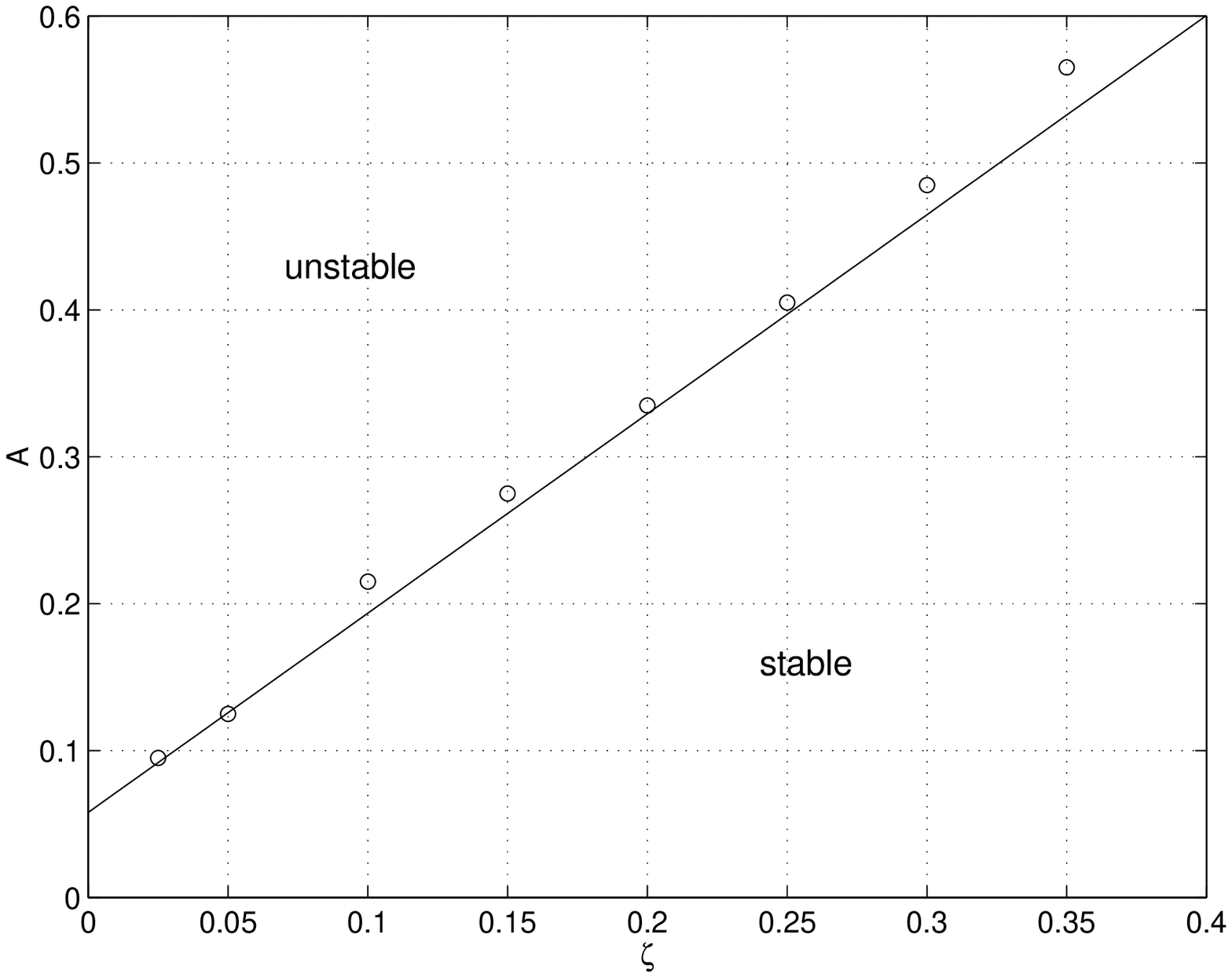,height=2.25in,width=2.25in}
\begin{center} (b) \end{center}
\end{minipage}
\end{center}
\vspace{0.2in}
\begin{center}
\begin{minipage}{2.25in}
\epsfig{figure=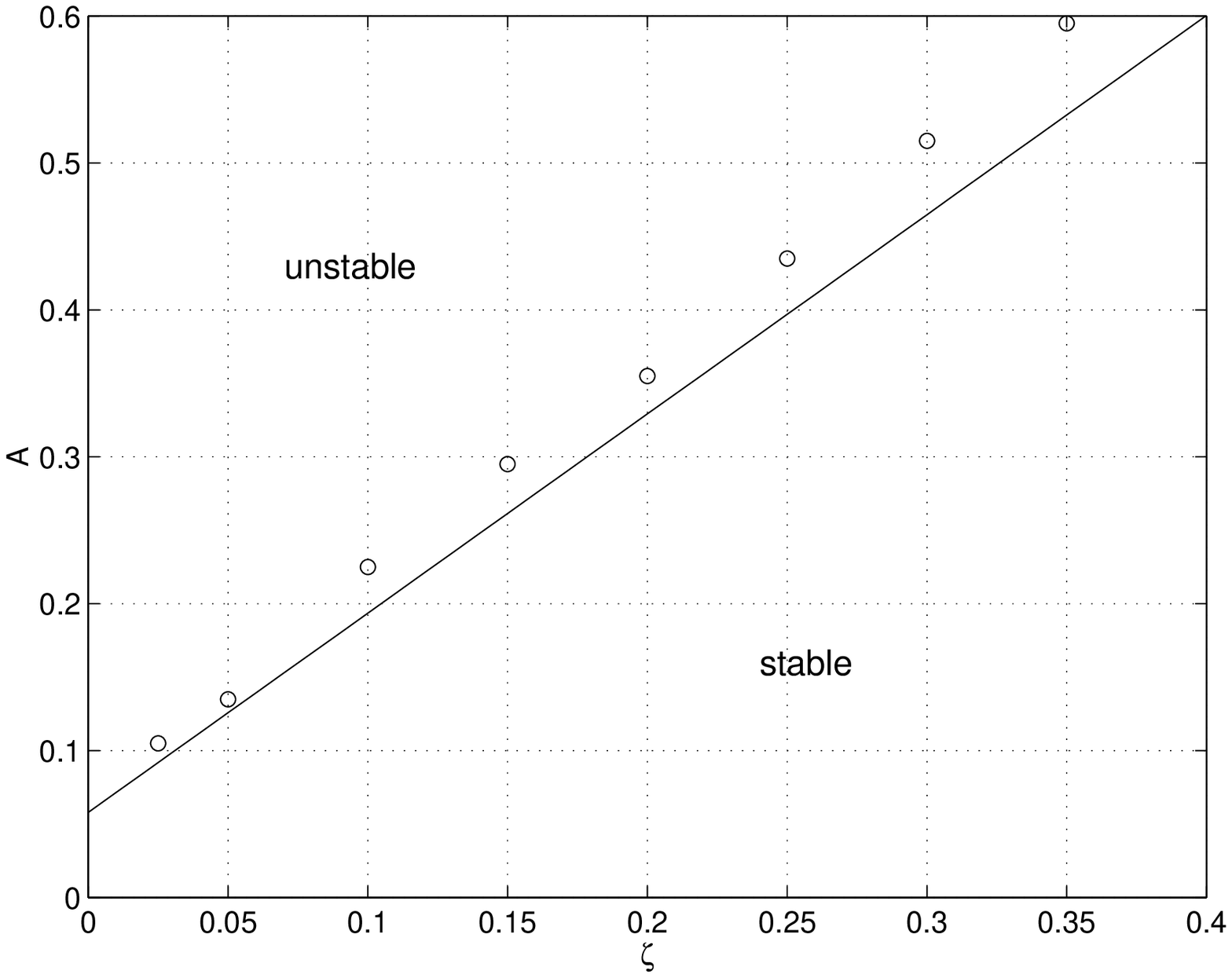,height=2.25in,width=2.25in}
\begin{center} (c) \end{center}
\end{minipage}
\hspace*{0.2in}
\begin{minipage}{2.25in}
\epsfig{figure=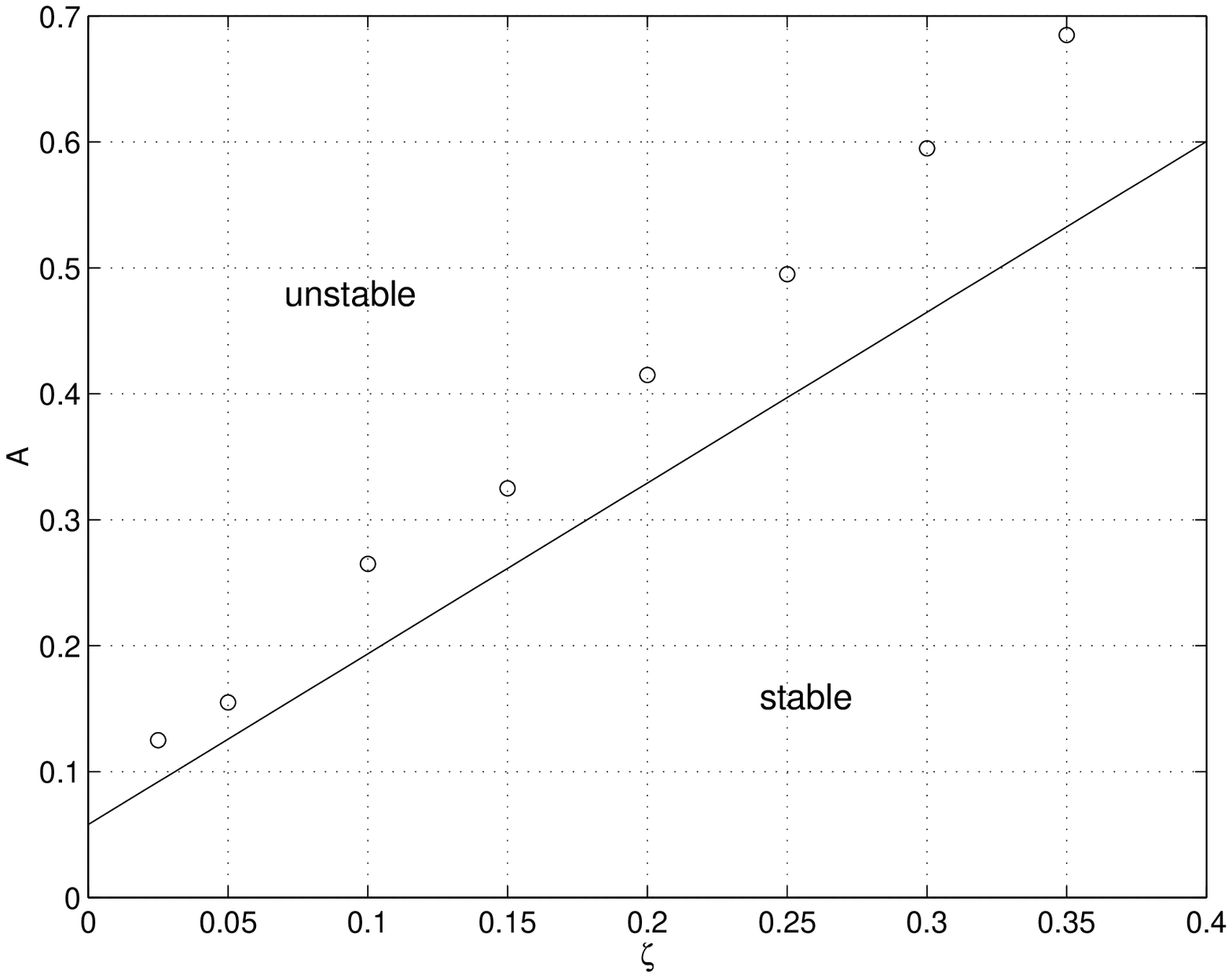,height=2.25in,width=2.25in}
\begin{center} (d) \end{center}
\end{minipage}
\end{center}
\caption{}
\label{figure8}
\end{figure}

The stability threshold predicted by equation (\ref{stab_thresh}) can
be compared to data obtained from simulating the Rayleigh-Plesset
equation (\ref{rp_eqn}) directly.  For small values of $\epsilon$,
figure \ref{figure8} shows the resulting good agreement.

The following is a brief description of how the simulations were
carried out. The material parameters used to produce figure
\ref{figure8} were
$\rho = 998$ kg/m$^{3}$,
$\mu = 0.001$ kg/m$\cdot$s, 
and
$\sigma = 0.0725$ N/m.
Four values of $\epsilon$ were chosen,
$\epsilon=0.01, 0.05, 0.1,$ and $0.2$.
For each fixed value of $\epsilon$ and for a selected set of values
of $\zeta$ ranging from 0 to 0.4, the parameters
$R_{0}$, $R_{\rm crit}$, $(p_{v}-p_{0}^{\infty})$, and $\Omega$
were calculated successively using the formulae:
$R_{0}=2 \mu^{2} / (\zeta^{2} \epsilon \rho \sigma )$,
$R_{\rm crit}=R_{0} / (1-\epsilon /2)$,
$p_{v}-p_{0}^{\infty}=(2 \sigma / R_{0})[1-(1/3)/(1- \epsilon /2)^{2}]$,
and
$\Omega = [(2 \sigma \epsilon)/(\rho R_{0}^{3})]^{1/2}$.
(Note that this succession of computations is done for each chosen
value of $\zeta$ in each of the four plots.)

Having obtained the dimensional parameters required for the simulation
of the full Rayleigh-Plesset equations corresponding to a given
($\epsilon$, $\zeta$) pair, we used a bisection procedure to determine
$A_{\rm esc}^{RP}$, the threshold value of $A$ separating bounded and
unbounded bubble trajectories.  The bisection procedure was initiated
by choosing a value of $A$ close to the linear regression line.  For
this choice of $A$, the dimensional pressure, $p_{A}$, was calculated
using the middle equation in (\ref{dist_lim_params}).  Then, the
initial condition $R(0)=R_{0}$ and $\dot{R}(0)=0$ was integrated
forward in time using an {\em implicit\/} \cite{numerical_recipes}
fourth-order Runge-Kutta scheme.  The adaptive, implicit scheme we used 
offers an accurate and stable means to integrate the governing equations.
The time steps are large in those intervals in which the bubble radius 
does not change rapidly, and they are extremely short for the intervals 
where $\dot{R}$ or $\ddot{R}$ is large (see for example figure 4.7 on 
page 309 of \cite{leighton}).  If the bubble radius remained
bounded during the simulation, then the value of $A$ was increased
slightly, a new $p_{A}$ was calculated, and a new simulation was
begun.  If, on the other hand, the bubble radius became unbounded
during the simulation, then the value of $A$ was slightly decreased
and a new simulation was initiated.  Continuing with this bisection of
$A$, the threshold value, $A_{\rm esc}^{RP}$, where the bubble first
becomes unstable was determined.

\begin{figure}[h!]
\begin{center}
\begin{minipage}{2.25in}
\epsfig{figure=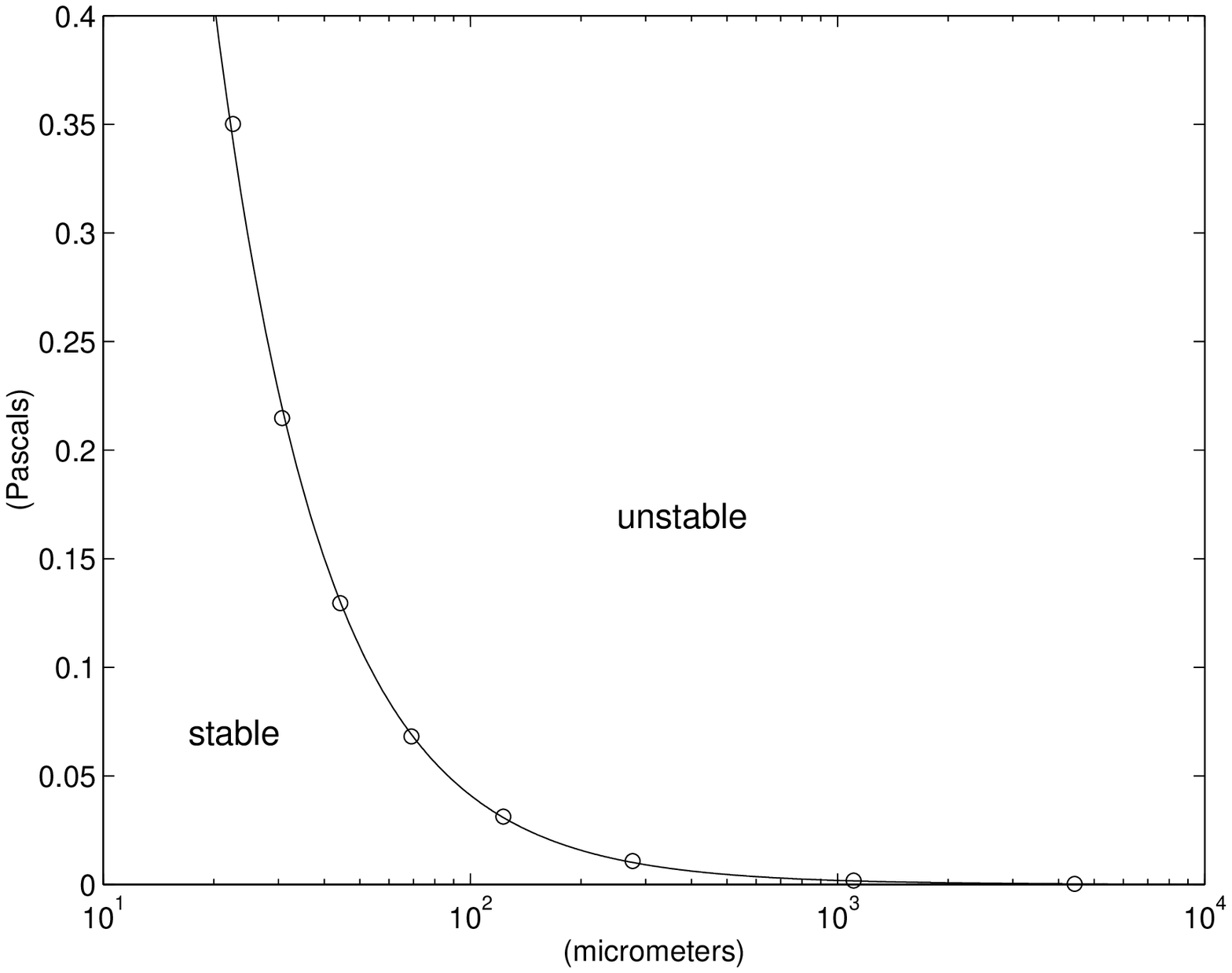,height=2.25in,width=2.25in}
\begin{center} (a) \end{center}
\end{minipage}
\hspace*{0.2in}
\begin{minipage}{2.25in}
\epsfig{figure=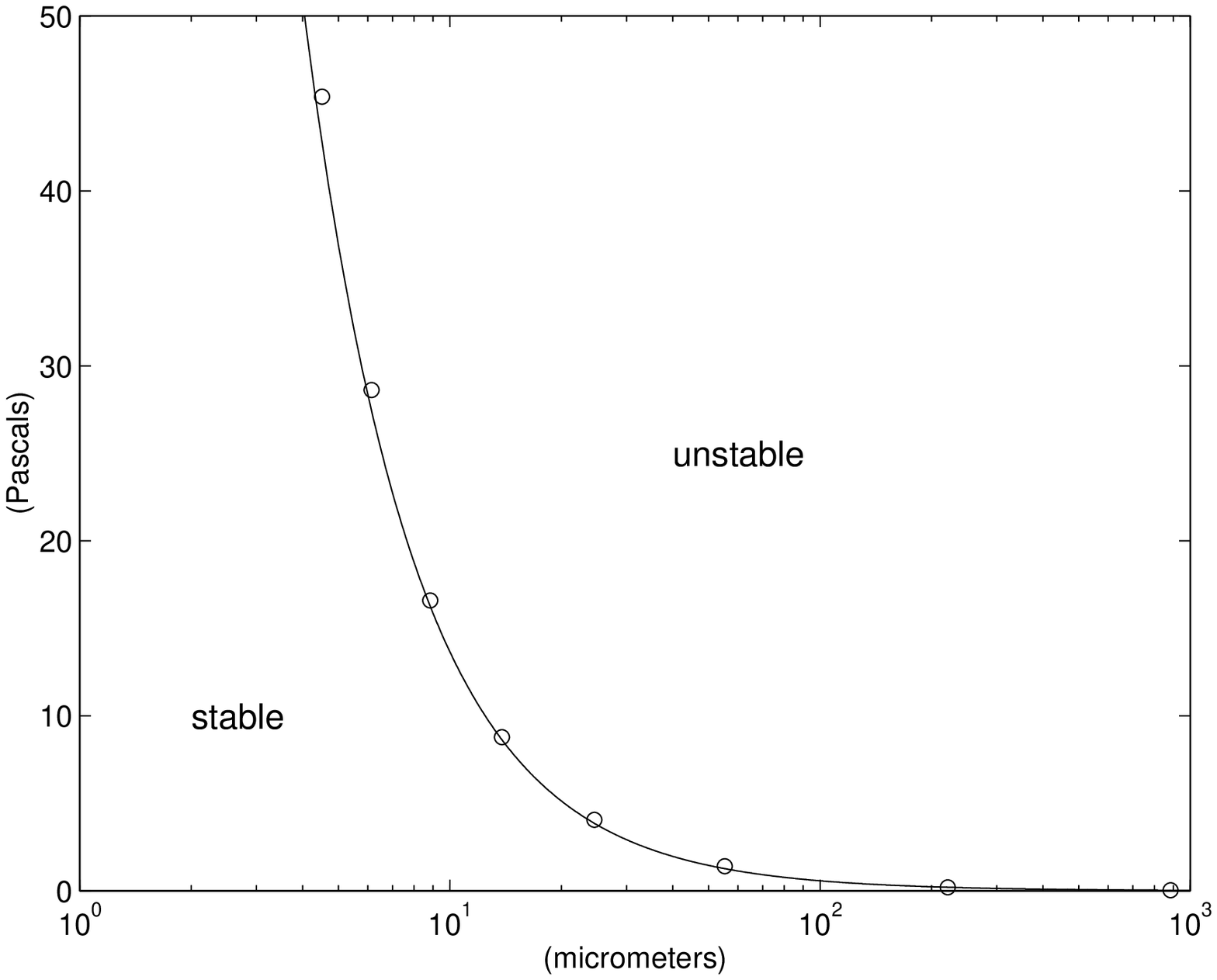,height=2.25in,width=2.25in}
\begin{center} (b) \end{center}
\end{minipage}
\end{center}
\vspace{0.2in}
\begin{center}
\begin{minipage}{2.25in}
\epsfig{figure=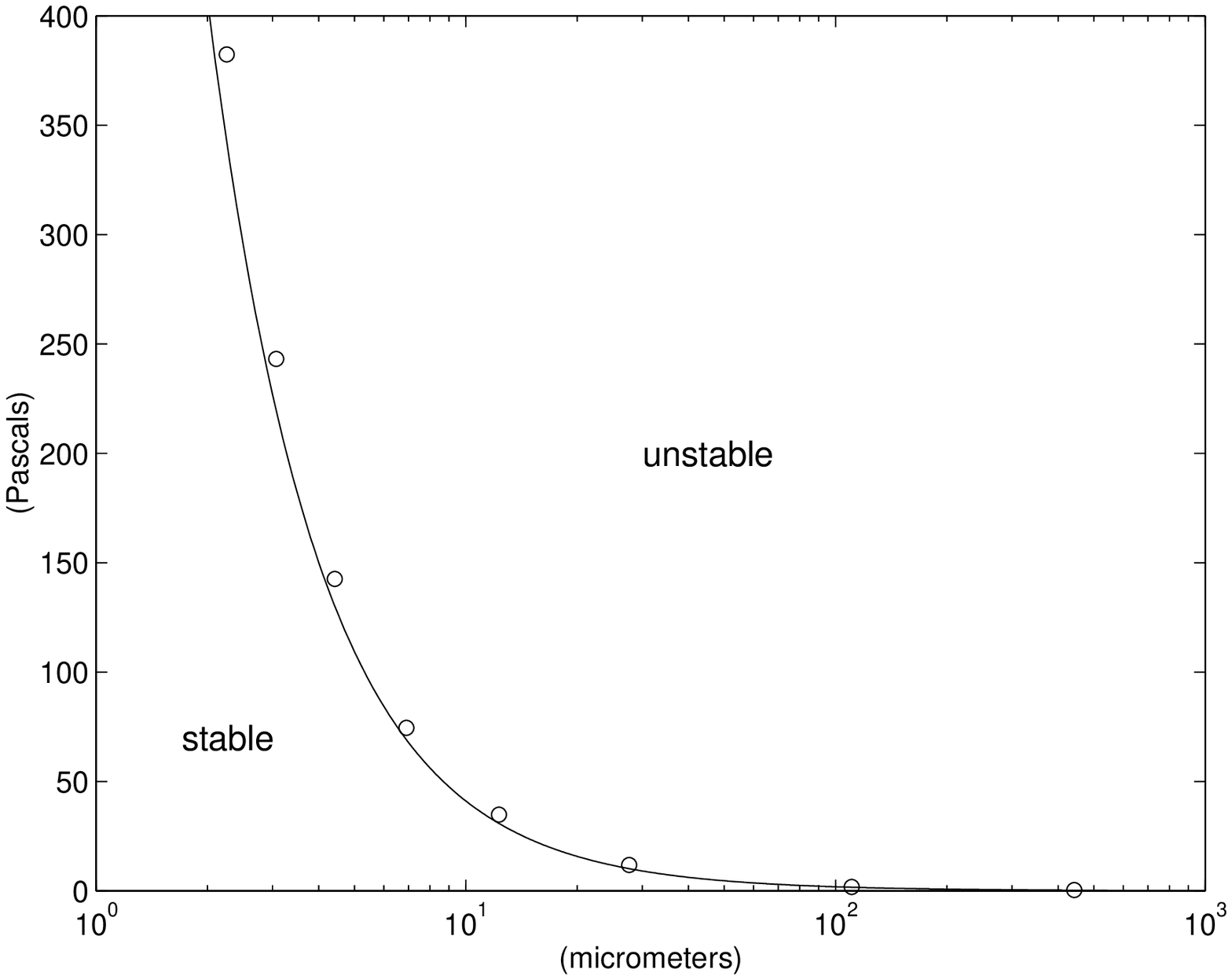,height=2.25in,width=2.25in}
\begin{center} (c) \end{center}
\end{minipage}
\hspace*{0.2in}
\begin{minipage}{2.25in}
\epsfig{figure=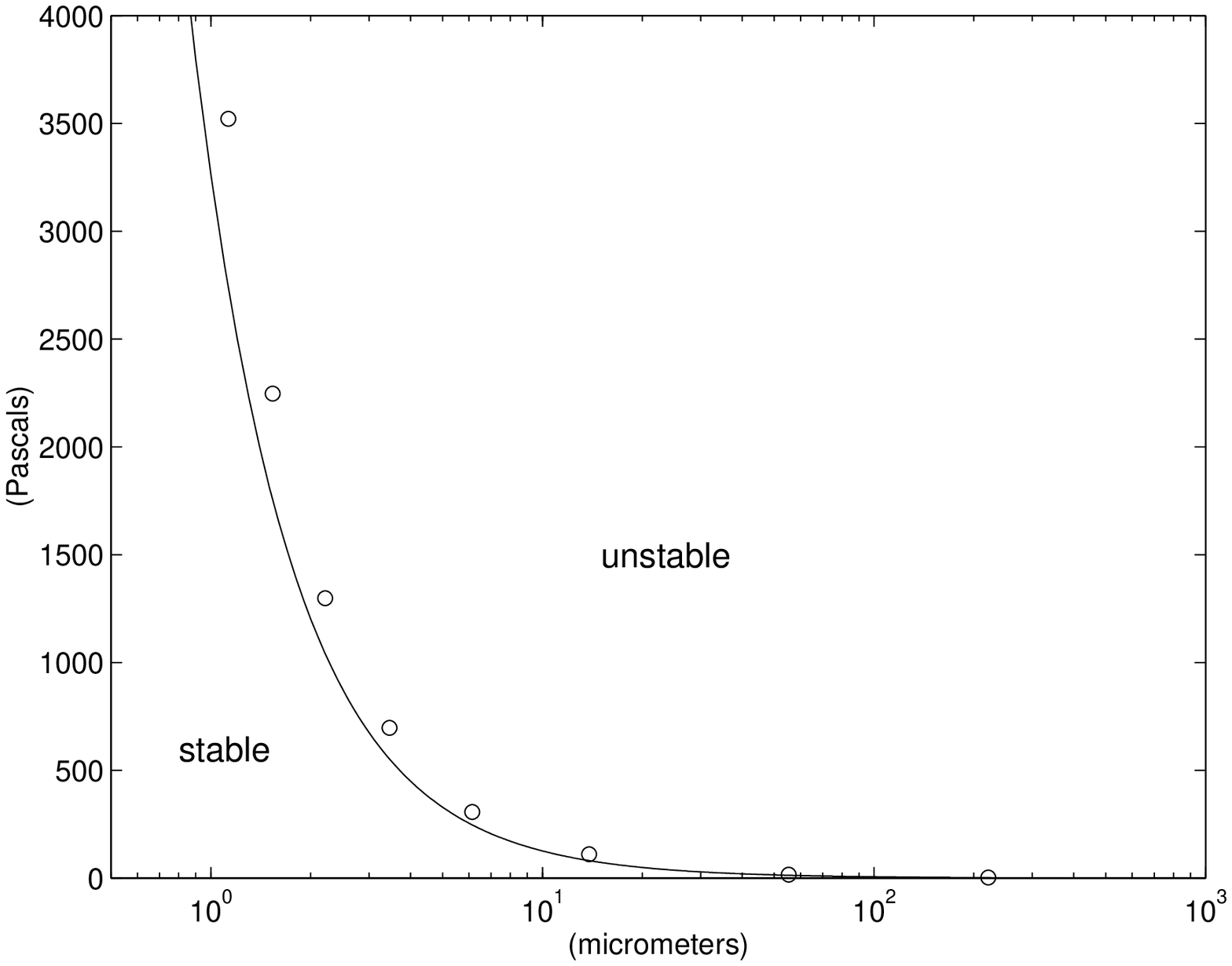,height=2.25in,width=2.25in}
\begin{center} (d) \end{center}
\end{minipage}
\end{center}
\caption{}
\label{figure9}
\end{figure}

The dimensional counterpart ($p_{A}$-versus-$R_{0}$) to figure
\ref{figure8} is shown in figure \ref{figure9} along with the
dimensional stability curve given by equation (\ref{stab_thresh}).
Note that for a given parameter $\epsilon$, the relationship which
defines the dimensionless damping parameter $\zeta$, {\em i.e.},
$R_{0}=2 \mu^{2} / (\zeta^{2} \epsilon \rho \sigma )$, can be thought
of as defining the bubble radius $R_0$. That is, for a given liquid
viscosity $\mu$ and with all other physical parameters being constant,
$\zeta$ can only change by varying the equilibrium radius $R_0$. As
such, the dimensionless $A$-versus-$\zeta$ curves can be put in terms
of the dimensional $p_A$-versus-$R_0$ curves, drawn in figure
\ref{figure9}.

\begin{figure}[h!]
\begin{center}
\epsfig{figure=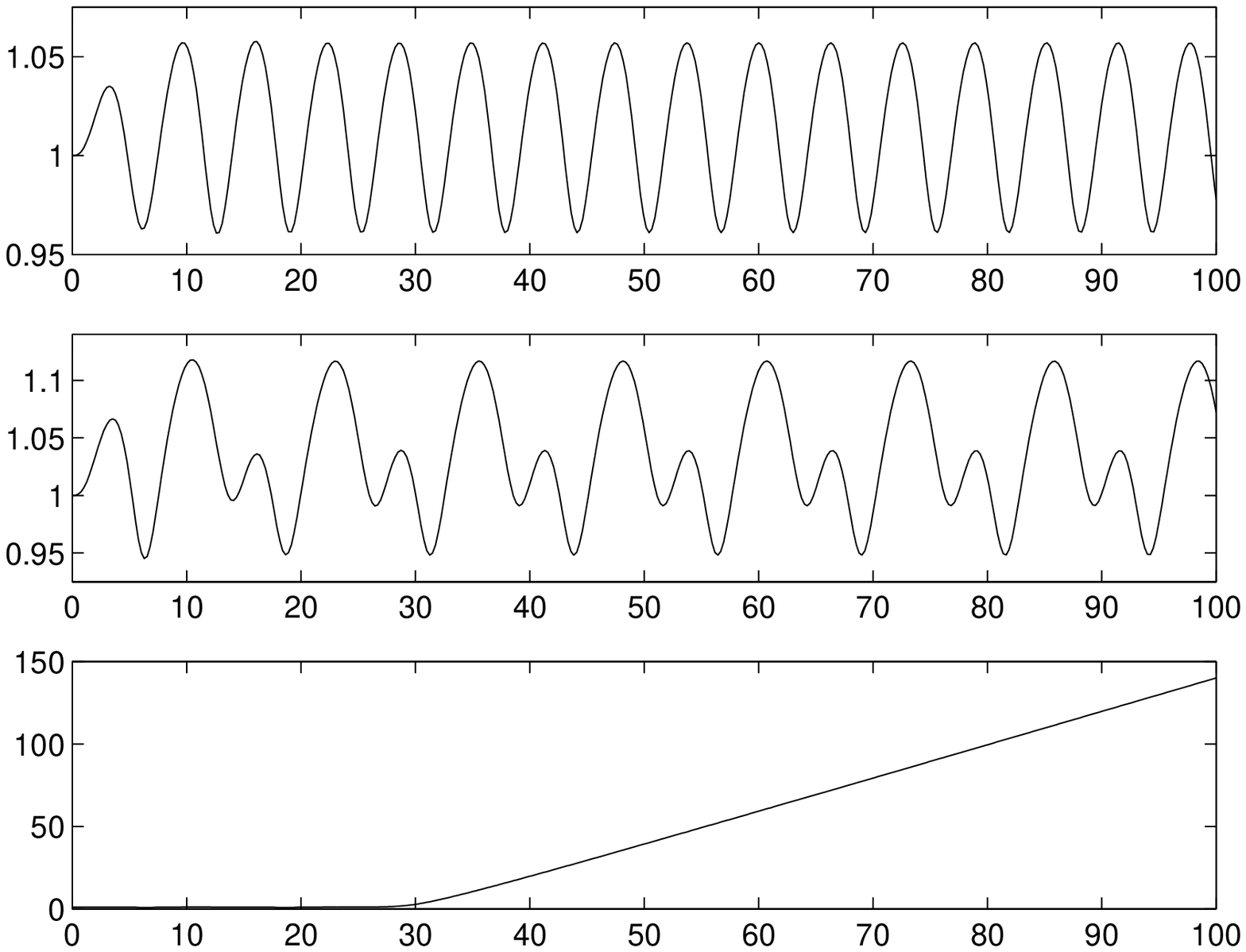,height=3in,width=6in}
\end{center}
\caption{}
\label{figure10}
\end{figure}

To show the way in which the bubble radius actually becomes unbounded
in the full Rayleigh-Plesset simulations, figure \ref{figure10}
provides the radius-versus-time plots for three typical simulations
with the same value of $\zeta=0.3$, where time is nondimensional.  In
this case, $A_{\rm esc}^{RP} \approx 0.51$. The top two curves are
obtained for values of $A$ of 0.3 and 0.5, respectively.  They show
stable oscillations although a period-doubling can be seen to have
occurred in going from one to the other. The bottom figure corresponds
to $A=0.53$ and shows that the bubble radius is becoming unbounded.
The corresponding dimensional parameters for the Rayleigh-Plesset
simulations are given in the figure caption.

\subsection{Consistency of the distinguished limit equation}
\label{Cotdle}

In this subsection, we argue, {\em a posteriori}, that it is
self-consistent to use the escape oscillator as the distinguished
limit equation (\ref{dist_lim_eqn}) for the full Rayleigh-Plesset
(\ref{rp_eqn}), {\em i.e.}, we show that the higher-order terms
encountered during the change of variables from $R$ to $x$ may be
neglected in a consistent fashion.

Recall that, in the derivation of the escape oscillator, all of the
${\cal O}(1)$ and ${\cal O}(\epsilon )$ terms dropped out, and the
ordinary differential equation (\ref{dist_lim_eqn}) was obtained by
equating the terms of ${\cal O}(\epsilon^{2})$.  The remaining terms
are of ${\cal O}(\epsilon^{3})$ and higher.  To be precise, at ${\cal
O}(\epsilon^{3})$, we find on the left-hand side:
\begin{displaymath}
x \ddot{x} - 2 \zeta x \dot{x} + \frac{3}{2} \dot{x}^{2},
\end{displaymath}
and on the right-hand side:
\begin{displaymath}
-\frac{3}{4}x + 2 x^{2} - \frac{20}{3} x^{3}.
\end{displaymath}
Moreover, we note that, for $i \geq 4$, all terms of ${\cal
O}(\epsilon^{i})$ on the left-hand side are of the form $x^{i-2}
\dot{x}$, while all terms of ${\cal O}(\epsilon^{i})$ on the
right-hand side are polynomials in $x$.

We already know that, for trajectories of the escape oscillator that
remain bounded, the $x$ and $\dot{x}$ variables stay ${\cal O}(1)$.
Hence, all of the higher-order terms remain higher-order for these
trajectories.  Next, for trajectories that eventually escape ({\em
i.e.}, those whose $x$-coordinate exceeds some large cut-off at some
finite time), we know that $x$ and $\dot{x}$ are bounded until that
time and afterwards they grow without bound.  The fact that then $R$
also grows without bound for these trajectories (due to the change of
variables that defines $x$) is consistent with the dynamics of the
full Rayleigh-Plesset equation.

Potential trouble could arise with trajectories for which $x$ becomes
negative and large in magnitude, {\em e.g.}, when $x \sim -1 /
\epsilon$ the coefficient of $\ddot{x}$ vanishes.  (This corresponds
to small R.)  A glance at the Poincar\'{e} map for the escape oscillator
reveals, however, that trajectories which have $x \sim -1 / \epsilon$
at some time $\tau$ can never have $x(\tau )$ and $\dot{x}(\tau )$ of
${\cal O}(1)$ simultaneously, for any $\tau$.  Hence, these
trajectories are not in the regime of interest, neither for the escape
oscillator nor for the full Rayleigh-Plesset equation.  This completes
the argument that it is self-consistent to use the escape oscillator
for this study.

\section{Pressure thresholds for general $\Omega^{*}$}
\label{Ptfgo}

\subsection{Stability curves for various $\Omega^{*}$}
\label{Scfvo}

Until now, we have only examined the case $\Omega^{*}=1$ in the
distinguished limit equation (\ref{dist_lim_eqn}).  In this section we
examine the dependence of the stability threshold on the acoustic
frequency, $\Omega$, for subcritical bubbles.  Specifically, for
frequencies $\Omega^{*}$ between 0.1 and 1.1, we performed numerical
simulations of the distinguished limit equation (\ref{dist_lim_eqn})
to determine many $(\zeta, A_{\rm esc})$ pairs.  These pairs are
plotted in figure \ref{figure11}, and the data points at each
dimensionless frequency $\Omega^*$ are connected by straight lines (in
contrast to the least squares fitting done in subsection \ref{Ppcfac}).

As in subsection \ref{Ppcfac}, good agreement between the distinguished 
limit equation threshold and the full Rayleigh-Plesset equation is 
observed for various values of $\Omega^{*}$; this can been seen in figure
\ref{figure12} which compares the two results at four different values
of $\Omega^{*}$ given by 0.6, 0.7, 0.8 and 0.9, for a fixed value of
$\epsilon=0.05$.

Various features observed in figure \ref{figure11}, such as the
flattening of these curves as $\Omega^{*}$ decreases, are explained in
the next subsection.

\subsection{Minimum forcing threshold}
\label{Mft}

Suppose that we wish to determine the driving frequency, for a given
bubble with an equilibrium radius $R_{0}$ and a critical radius
$R_{\rm crit}$, so that the acoustic forcing amplitude necessary to 
make the bubble unstable is minimized.
This can be done by choosing in figure
\ref{figure11}, the value of $\Omega^{*}$ for which the corresponding
threshold curve is below all the others for a given $\zeta$. The
result of such a procedure is provided in figure \ref{figure13} as
follows.  Figure \ref{figure13}(a) provides the frequency of harmonic
forcing at a given value of the damping parameter $\zeta$ for which
the required amplitude of the acoustic pressure field to create
cavitation is the smallest. Figure \ref{figure13}(b) shows the
dimensionless minimum pressure amplitude $A_{\rm esc}$ corresponding
to the value of $\Omega^{*}$ just presented. In figure
\ref{figure13}(a), for $\zeta$ between 0 and 0.225, the frequency
curve is nearly a straight line and we fit a linear regression line to
the data in that interval: $\Omega^{*} = -1.12 \zeta + 0.90$ for $0
\le \zeta \le 0.225$.  Correspondingly, in figure \ref{figure13}(b),
we see that the minimum pressure curve is also nearly straight for the
same interval of $\zeta$ values.  The least squares line fitting the
data in figure \ref{figure13}(b) is $A = 1.03 \zeta + 0.02$ for $0 \le
\zeta \le 0.225$.

When $\zeta=0.225$, there is a discontinuity in the frequency curve,
as seen in figure \ref{figure13}(a).  At the same value of $\zeta$,
the pressure curve levels off to $A \approx 0.25$.  This can be
explained by a brief analysis of the normal form equation.  The key
observation will be that, in the escape oscillator with constant
forcing ({\em i.e.}, constant right-hand side), there is a saddle-node
bifurcation when the magnitude of the forcing is 1/4.  In order to
carry out this brief analysis, we consider the cases $\zeta>0$ and
$\zeta=0$ separately, beginning with $\zeta>0$.

For $\zeta>0$, the Poincar\'{e} map of the normal form equation
(\ref{dist_lim_eqn}) has an asymptotically stable fixed point (a sink), 
which corresponds to an attracting periodic orbit for the full normal
form equation.  Now, during each period of the external forcing, the
location of this periodic orbit in the $(x,\dot{x})$-plane changes.
In fact, for the small values of $\Omega^{*}$ we are interested in
here ($\Omega^{*} \le 0.3$ approximately), the change in location occurs
slowly, and one can write down a perturbation expansion for its position
in powers of the small parameter $\Omega^{*}$.  The coefficients at each
order are functions of the slow time $z \equiv \Omega^{*} t$.  To leading
order, {\em i.e.,} at ${\cal O}(1)$, the attracting periodic orbit is
located at the point $(x(z),0)$, where $x(z)$ is the smaller root of
$x-x^{2} = A \sin (z)$, namely
\begin{displaymath}
x(z)=\frac{1}{2} - \frac{1}{2}\sqrt{1 - 4 A \sin(z)}.
\end{displaymath}

Therefore, one sees directly that $A=1/4$ is a critical value. In
particular, if one considers any fixed value of $A< 1/4$, then the
attracting periodic orbit exists for all time, and the trajectory of
our initial condition $(0,0)$ will be always be attracted to it.
(Note that the viscosity $\zeta \ge 0.225$ is large enough so orbits
are attracted to the stable periodic orbit at a faster rate than the
rate at which the periodic orbit's position moves in the
$(x,\dot{x})$-plane due to the slow modulation.)  However, for any
fixed value of $A>1/4$, the function giving $x(z)$ becomes complex
after the slow time $z$ reaches a critical value $z_{*}(A)$, where $A
\sin (z_{*}) = 1/4$, and where we write $z_{*}(A)$ since $z_{*}$
depends on $A$.  Moreover, $x(z)$ remains complex in the interval
$(z_{*}(A),\pi -z_{*}(A))$ during which $A \sin (z) > 1/4$.  Viewed in
terms of the slowly-varying phase portrait, the slowly-moving sink
merges with the slowly-moving saddle in a saddle-node bifurcation when
$z$ reaches $z_{*}(A)$, and they disappear together for $z_{*}(A) < z
< \pi -z_{*}(A)$.  Hence, the attracting periodic orbit no longer
exists when $z$ reaches $z_{*}(A)$, and the trajectory that started at
$(0,0)$ --- and that was spiraling in toward the slowly-moving
attracting periodic orbit while $z$ was less than $z_{*}(A)$ ---
escapes, because there is no longer any attractor to which it is
drawn.

For the sake of completeness in presenting this analysis, we note that
when $A=1/4$, then $z_{*}(A)=\pi /2$; hence, it is precisely near this
lowest value of $A$, namely $A=1/4$, that we find the threshold for the
acoustic forcing amplitude, and the escape happens near the slow time
$z=\pi /2$.  Moreover, for values of $A> 1/4$, $0< z_{*}(A) < \pi /2$,
and so the escape happens at an earlier time.

Numerically, the minimal frequency $\Omega^{*}$ appears to be
$\Omega^{*} \approx 0$, where $\Omega^{*}=0.01$ is the lowest value for
which we conducted simulations.  Moreover, this also explains why,
as we see from figure \ref{figure11} already, the curves are flat with
$A \approx 0.25$ for $\Omega^{*} \le 0.3$.  This is the range of small
values of $\Omega^{*}$ for which the above analysis applies.

Next, having analyzed the regime in which $\zeta >0$,
we turn briefly to the case $\zeta=0$.  For small values,
$\Omega^{*} \le 0.3$, the curves in figure \ref{figure11} remain flat
near $A=0.25$ all the way down to $\zeta=0$.  The full normal form equation
with $\zeta=0$ is a slowly-modulated Hamiltonian system.  One can again use
the slowly-varying phase planes as a guide to the analysis (although the
periodic orbit is only neutrally stable when $\zeta=0$ and no longer
attracting as above), and the saddle-node bifurcation in the leading
order problem at $A \sin (z_{*})=1/4$ is the main phenomenon responsible
for the observation that the threshold forcing amplitude is near 0.25.
(We also note that for a detailed analysis of the trapped orbits, one
needs adiabatic separatrix-crossing theory, see
\cite{cary}, for example, but we shall not need that here.)

Finally, and most importantly, simulations of the full
Rayleigh-Plesset equation confirm all the quantitative features of
this analysis of the normal form equation.  The open circles in figure
\ref{figure13} represent the numerically observed threshold forcing
amplitudes, and these circles lie very close to the curves obtained as
predictions from the normal form equation.  We attribute this
similarity to the fact that the phase portrait of the isothermal
Rayleigh-Plesset equation has the same structure --- stable and
unstable equilibria, separatrix bounding the stable oscillations, and
a saddle-node bifurcation when the forcing amplitude exceeds the
threshold --- as the normal form equation (see \cite{szeri-leal}).

\subsection{The dimensional form of the minimum forcing threshold}
\label{Tdfotmft}

Recall, that for $\zeta$ between 0 and 0.225, we fit linear regression
lines to portions of figures \ref{figure13}(a) and (b).  Specifically,
for figure \ref{figure13}(a) we found that, for a particular choice of
$\zeta$, the frequency which yields the smallest value of $A_{\rm
esc}$ can be expressed as: $\Omega^{*} = -1.12 \zeta + 0.90$ for $0
\le \zeta \le 0.225$.  And for figure \ref{figure13}(b) we found that
the stability boundary for the minimum forcing is given by,
\begin{equation}
A = \left\{
\begin{array}{lll}
1.03 \zeta + 0.02 & \mbox{for} & 0 \le \zeta \le 0.225 \\
0.25              & \mbox{for} & 0.225 \le \zeta \le 0.4\,.
\end{array}
\right. 
\label{optimal_nondim_thresh}
\end{equation}
Using the definitions of $\zeta , A$ and $\Omega^{*}$ as given by
(\ref{dist_lim_params}), the ``optimal'' acoustic frequency to cause
cavitation of a subcritical bubble is given in dimensional form by
\begin{displaymath}
\Omega =
-2.24 \frac{\mu}{\rho R_{0}^{2}} +
1.27 \frac{\epsilon^{1/2} \sigma^{1/2}}{\rho^{1/2} R_{0}^{3/2}} 
\hspace{1.5em} \mbox{for} \hspace{1.5em}
0 \le \left( 
\frac{2 \mu^{2}}{\sigma \epsilon \rho R_{0}}
\right)^{1/2} \le 0.225 \,.
\end{displaymath}
Correspondingly, the minimum acoustic pressure threshold is
\begin{displaymath}
p_{A} > \left\{
\begin{array}{lll}
2.91 \frac{\epsilon^{3/2} \sigma^{1/2} \mu}{\rho^{1/2} R_{0}^{3/2}} +
0.04 \frac{\epsilon^{2} \sigma}{R_{0}}
& \mbox{for} & 0 \le \left( 
\frac{2 \mu^{2}}{\epsilon \sigma \rho R_{0}}
\right)^{1/2} \le 0.225 \\
\frac{\epsilon^{2} \sigma}{2 R_{0}}
& \mbox{for} & 0.225 \le \left(
\frac{2 \mu^{2}}{\epsilon \sigma \rho R_{0}}
\right)^{1/2} \le 0.4 \,.
\end{array}
\right. 
\end{displaymath}

\subsection{A lower bound for $A_{\rm esc}$ via Melnikov analysis}
\label{Albfavma}

The distinguished limit equation (\ref{dist_lim_eqn}) can be
written as the perturbed system
\begin{displaymath}
\mathbf{\dot{x} = f(x)} + \tilde{\epsilon} \mathbf{g(x}, \tau)
%\bf{ \dot{x} = f(x) + \tilde{\epsilon} {g(x,} \tau) }
\end{displaymath}
where, $\mathbf{x}=(x,y)$,
$\mathbf{f}(x,y) = (y,x^{2}-x)$
%$\bf{f}(x,y) = (y,x^{2}-x)$
and
$\mathbf{g}(x,y,\tau) =
%$\bf{g}(x,y,\tau) =
\left( 0, \bar{A} \sin (\Omega^{*} \tau ) - 2 \bar{\zeta} y \right)$
with
$A= \tilde{\epsilon} \bar{A}$, $\zeta= \tilde{\epsilon} \bar{\zeta}$.
When $\tilde{\epsilon} =0$ the system has a center at (0,0) and a
saddle point at (1,0).  The homoclinic orbit to the unperturbed saddle
is given by
$\gamma_{0}(\tau) = (x(\tau),y(\tau))$ where
$x(\tau) = -(1/2) + (3/2) \tanh^{2} (\tau /2)$
and
$y(\tau) = (3/2) \tanh (\tau /2) {\rm sech}^{2} (\tau /2)$.
Following \cite{guckenheimer-holmes}, the Melnikov function takes the form
\begin{eqnarray*}
M(\tau_{0}) & = & \int_{- \infty}^{\infty}
\mathbf{f}(\gamma_{0}(\tau)) \wedge
\mathbf{g}(\gamma_{0}(\tau),\tau + \tau_{0})
%\bf{f}(\gamma_{0}(\tau)) \wedge
%\bf{g}(\gamma_{0}(\tau),\tau + \tau_{0})
\, d \tau
\\
& = & \frac{3}{2} \bar{A} \int_{- \infty}^{\infty}
\sin[\Omega^{*} (\tau + \tau_{0})]
\tanh \left( \frac{\tau}{2} \right)
{\rm sech}^{2} \left( \frac{\tau}{2} \right)
\, d \tau
\\
& & \hspace*{0.2in}
- \; \frac{9}{2} \bar{\zeta} \int_{- \infty}^{\infty}
\tanh^{2} \left( \frac{\tau}{2} \right)
{\rm sech}^{4} \left( \frac{\tau}{2} \right)
\, d \tau.
\end{eqnarray*}
The first integral can be done with a residue calculation, 
and the second integral is evaluated in a straightforward manner,
resulting in:
\begin{displaymath}
M(\tau_{0}) =
-\left[
\frac{6 \pi (\Omega^{*})^{2} \cos (\Omega^{*} \tau_{0})}
{\sinh (\pi \Omega^{*})}
\right]
\bar{A}
- \frac{12}{5} \bar{\zeta}.
\end{displaymath}

The Melnikov function has simple zeros when
$\bar{A} > \bar{A}_{\rm h.tan.}$, where
\begin{equation}
\bar{A}_{\rm h.tan.} =
\left(
\frac{2 \sinh (\pi \Omega^{*})}{5 \pi (\Omega^{*})^{2}}
\right)
\bar{\zeta}.
\label{homotan}
\end{equation}
Hence the stable and unstable manifolds of the perturbed saddle
point intersect transversely for all sufficiently small
$\tilde{\epsilon} \neq 0$ when $\bar{A} > \bar{A}_{\rm h.tan.}$
\cite{guckenheimer-holmes}. The resulting chaotic dynamics is evident 
in figure \ref{figure5}, for example.  Since homoclinic tangency must 
occur before the trajectory through the origin can escape,
$\bar{A}_{\rm h.tan.}$ may be viewed as a precursor to
$A_{\rm esc}$.  Figure \ref{figure14} demonstrates that, for small
enough $\tilde{\epsilon}$, equation (\ref{homotan}) provides a lower
bound for the stability curves seen in figure \ref{figure11}.

The reason why Melnikov analysis yields a lower bound for the cavitation
threshold relates to how deeply the stable and unstable manifolds of the
saddle fixed point of the Poincare map for equation (\ref{dist_lim_eqn})
penetrate into the region bounded by the separatrix in
the $A, \zeta=0$ case.  For sufficiently small values of $\tilde{\epsilon}$, 
long segments of the perturbed local stable and unstable manifolds will stay 
${\cal O}(\tilde{\epsilon})$ close to the unperturbed homoclinic orbit.
However, as $\tilde{\epsilon}$ grows (and one gets out of the regime in 
which the asymptotic Melnikov theory strictly applies), these local manifolds 
will penetrate more deeply into the region bounded by the separatrix in
the $A, \zeta=0$ case.
In fact, there is a sizable gap in the parameter space between
the homoclinic tangency values and the escape values corresponding to
our cavitation criteria, {\it i.e.,} when the trajectory through the origin
grows without bound. There is a similar gap when other initial
conditions are chosen.

The Melnikov function was also calculated in \cite{thompson}.
There, a detailed analysis of escape from a cubic potential is described
and the fractal basin boundaries and occurrence of homoclinic tangencies
are given.
We also refer the reader to \cite{grimshaw-tian} in which a closely related 
second order, damped and driven oscillator with quadratic nonlinearity
is studied using both homoclinic Melnikov theory, as was done here, and
subharmonic Melnikov theory.
The existence of periodic orbits is demonstrated there, and period doubling
bifurcations of these periodic orbits are examined.
Their equation arises from the study of travelling waves in a forced, damped
KdV equation.

\section{Pressure fields with two fast frequencies}
\label{Pfwtff}

In this section, we consider what happens to the cavitation threshold if two 
fast frequency components are present in the acoustic pressure field,
and the slow transducer, which lowers the ambient pressure and whose effect 
is quasistatic, is also still present.
In figure \ref{figure15}, we show the results from simulations
with quasiperiodic pressure fields.  These were obtained from simulations
of (\ref{dist_lim_eqn}) with the forcing replaced by
$(A/2)(\sin (\Omega_{1}^{*} \tau ) + \sin (\Omega_{2}^{*} \tau ) )$,
and a wide range of values for $\Omega_{1}^{*}$ and $\Omega_{2}^{*}$.
For a fixed value of $\zeta=0.25$, the cavitation surface shown in the
figure was plotted by computing the triples
$(\Omega_{1}^{*},\Omega_{2}^{*},A_{\rm esc})$.

We note that in figure \ref{figure15}, the intersection of the cavitation 
surface and the vertical plane given by $\Omega_{1}^{*} = \Omega_{2}^{*}$ 
represents cavitation thresholds for acoustic forcing of the form 
$A\sin (\Omega^{*} \tau )$ ({\it i.e.,} 
a single fast frequency component and a 
quasistatic component).  Furthermore, we see that the global minimum of the 
cavitation surface lies along the line $\Omega_{1}^{*} = \Omega_{2}^{*}$.
Hence, for $A/2$ as our particular choice of quasiperiodic forcing 
coefficient, the addition of a second fast frequency component in the 
pressure field does not lower the cavitation threshold beyond that of
the single fast frequency case.

\section{Discussion}
\label{Discussion}

A distinguished limit equation has been derived which is suitable for
use in determining cavitation events of slightly subcritical bubbles.
This ``normal form" equation allows us to study cavitation thresholds
for a range of acoustic forcing frequencies.  For $\Omega^{*}=1$, we
find an explicit expression for the cavitation threshold via linear
regression, since the simulation data reveal an approximate linear
dependence of the nondimensional threshold amplitude, $A$, on the
nondimensional liquid viscosity, $\zeta$.  When converted to
dimensional form, this linear expression translates into a nonlinear
dependence, cf.\ equation (\ref{stab_thresh}), on the material
parameters.  In all of our simulations, the acoustic threshold
amplitude coincides with the amplitudes at which the cascades of
period-doubling subharmonics terminate.

Particular attention has also been paid to calculating the frequency,
$\Omega^{*}$, at which a given subcritical bubble will most easily
cavitate.  Expression (\ref{optimal_nondim_thresh}) for the
corresponding minimum threshold amplitude $A$ grows linearly in
$\zeta$ for $\zeta < 0.225$ until the critical amplitude $A=1/4$ is
reached, and the threshold amplitude stays constant at $A \approx 1/4$
for larger $\zeta$.  For these larger values of $\zeta > 0.225$, the
``optimal'' frequency is essentially zero, as we showed by doing a
slowly-varying phase portrait analysis and exploiting the fact that
the normal form equation undergoes a saddle-node bifurcation at
$A=1/4$ in which the entire region of bounded stable orbits vanishes.
The full Rayleigh-Plesset equation undergoes a similar bifurcation at
forcing amplitudes very near $A=1/4$ for sufficiently small
$\epsilon$.  Overall, the results from the normal form equation are in
excellent agreement with those of the full Rayleigh-Plesset equation,
and this may be attributed to the high level of similarity between the
phase-space structures of both equations.

In view of the findings in \cite{szeri-leal}, we may draw an additional 
conclusion from the present work.  In a certain sense, we have extended 
the finding of lowered transition amplitudes reported in \cite{szeri-leal} 
to the limiting case of one low frequency and one fast frequency. 
We find that if a low frequency transducer prepares a bubble to
become slightly subcritical, then the presence of a high frequency
transducer can lower the cavitation threshold of the bubble below
the Blake threshold.

Our results on the optimum forcing frequency and minimum pressure
threshold to cavitate a subcritical bubble may also be useful in
fine-tuning experimental work on single-bubble sonoluminescence
(SBSL). In SBSL \cite{gaitan,crum94,putterman}, a single bubble is
acoustically forced to undergo repeated cavitation/collapse cycles, in
each of which a short-lived flash of light is produced. While the
process through which a collapsing bubble emits light is very complex
and involves many nonlinear phenomena, the possibility of better
control over cavitation and collapse, {\em e.g.}, through the use of
multiple-frequency forcing, can perhaps be investigated using the type
of analysis presented in this paper.

\vspace*{7mm}

{\bf Acknowledgments} --- 
We are grateful to Professor S. Madanshetty for many helpful discussions.
The authors would also like to thank the referees for their comments.
This research was made possible by Group Infrastructure Grant DMS-9631755
from the National Science Foundation.
A.H. gratefully acknowledges financial support from the
National Science Foundation via this grant.
T.K. gratefully acknowledges support from the Alfred P. Sloan Foundation
in the form of a Sloan Research Fellowship.

\vspace*{7mm}

\section*{Appendix: Coaxing experiments in acoustic microcavitation}

To illustrate one application of our results, we now briefly
consider the experimental findings of \cite{madanshetty}
on so-called ``coaxing'' of acoustic microcavitation. In these
experiments, smooth submicrometer spheres were added to clean water
and were found to facilitate the nucleation of cavitation events ({\em
i.e.} reduce the cavitation threshold) when a high-frequency
transducer (originally aimed as a detector) was turned on at a
relatively low pressure amplitude. Specifically, the main cavitation
transducer was operating at a frequency of 0.75 MHz, while the active
detector had a frequency of 30 MHz. In a typical experiment, with
0.984-micron spheres added to clean water, the cavitation threshold in
the absence of the active detector was found to be about 15 bar peak
negative.  When the active detector was turned on, producing a minimum
pressure of only 0.5 bar peak negative by itself, it caused the
cavitation threshold of the main transducer to be reduced from 15 to 7
bar peak negative. The polystyrene latex spheres were observed under
scanning electron microscopes and their surface was determined to be
smooth to about 50 nanometers. It was thus thought that any gas
pockets which were trapped on their surface due to incomplete wetting
and which served as nucleation sites for cavitation, were smaller in
size than this length.  In \cite{madanshetty}, it is conjectured
that the extremely high fluid accelerations created by the
high-frequency active detector, coupled with the density mismatch
between the gas and the liquid, caused these gas pockets to accumulate
on the surface of the spheres and form much larger ``gas caps'' (on
the order of the particle size), which then cavitated at the lower
threshold. Here we attempt to provide an alternative explanation for
the observed lowering of the threshold in the presence of the active
detector.

To effect our estimates, we shall use the same physical parameters as
earlier: $\mu=0.001$ kg/m$\cdot$s, $\rho=0.998$ kg/m$^3$ and
$\sigma=0.0725$ N/m. We also ignore the vapor pressure of the liquid
at room temperature. Note also that 1 bar=$10^5$ N/m$^2$ and that
the transducer frequencies $f$ cited above are related to the radian
frequencies $\Omega$ used earlier by $\Omega=2\pi f$.

Let us begin by estimating a typical size for the nucleation sites
which cavitate at $p_{L_{\rm crit}}=-15$ bar in the absence of the
active detector. Upon using Blake's classical estimate of $p_{L_{\rm
crit}}= - 0.77 \sigma/R_0$, the equilibrium radius of the trapped air
pockets is estimated to be $R_0=3.7 \times 10^{-8}$ m or $37$ nm. This
size is consistent with the observation that the surfaces of the
spheres were smooth to within $50$ nm. We note that such a small
cavitation nucleus cannot exist within the homogeneous liquid itself
since it would dissolve away extremely fast due to its overpressure
resulting from surface tension. However, when trapped in a crevice or
within the roughness on solid surfaces, it can be stabilized against
dissolution with the aid of the meniscus shape which separates it from
the liquid. The natural frequency of a 37 nm bubble (if it were
spherical) found from equation (\ref{natural_freq}) would be 385 MHz
which is very large compared to forcing frequency of the cavitation
transducer which is 0.75 MHz. Therefore, consistent with Blake's
classical criterion, the pressure changes in the liquid would appear
quasistatic to the bubble and at such a small size, surface tension
does dominate the bubble dynamics. The Blake critical radius $R_{\rm
crit}$ which corresponds to this equilibrium radius $R_0$ of 37 nm
can be calculated to be $R_{\rm crit}=64$ nm.

Let us now suppose that the cavitation transducer is operating at
$p_L=-7$ bar peak negative as in the experiments with the active
transducer also turned on. Using equation (\ref{liquid_pressure_iso}),
the final expanded radius of the bubble when the liquid pressure is
quasistatically reduced to $-7$ bar is found to be $4.2 \times 10^{-8}$
m or $42$ nm. In other words, a bubble of original radius 37 nm at a
liquid pressure of 1 bar, grows to a maximum size of 42 nm when the
liquid pressure is reduced to $-7$ bar.  Its critical radius is still
$64$ nm, reached if the liquid pressure were to be reduced further to
$-15$ bar.

At this point, since the pressure changes in the liquid due to the
$0.75$ MHz cavitation transducer are occurring slowly compared both
with the natural timescale of the bubble and the 30 MHz detector, let
us take the mean pressure in the liquid to be the $p_0^\infty=-7$ bar,
and imagine the bubble size at this pressure to be its new equilibrium
radius $R_0=42$ nm, with the critical radius still given by $R_{\rm
crit}=64$ nm. This bubble is now assumed to be forced by the 30 MHz
transducer at an acoustic pressure amplitude of $1.5$ bar ({\em i.e.}
$-0.5$ bar peak negative).  Using these values, the perturbation
parameter $\epsilon$ is calculated from equation (\ref{eps}) to be
$\epsilon=0.69$. This parameter is too big for the results of the
asymptotic theory to provide meaningful quantitative agreement;
nevertheless, we proceed with the discussion to see if we can at least
obtain the right order of magnitude for the pressure threshold.

\begin{figure}[h!]
\begin{center}
\epsfig{figure=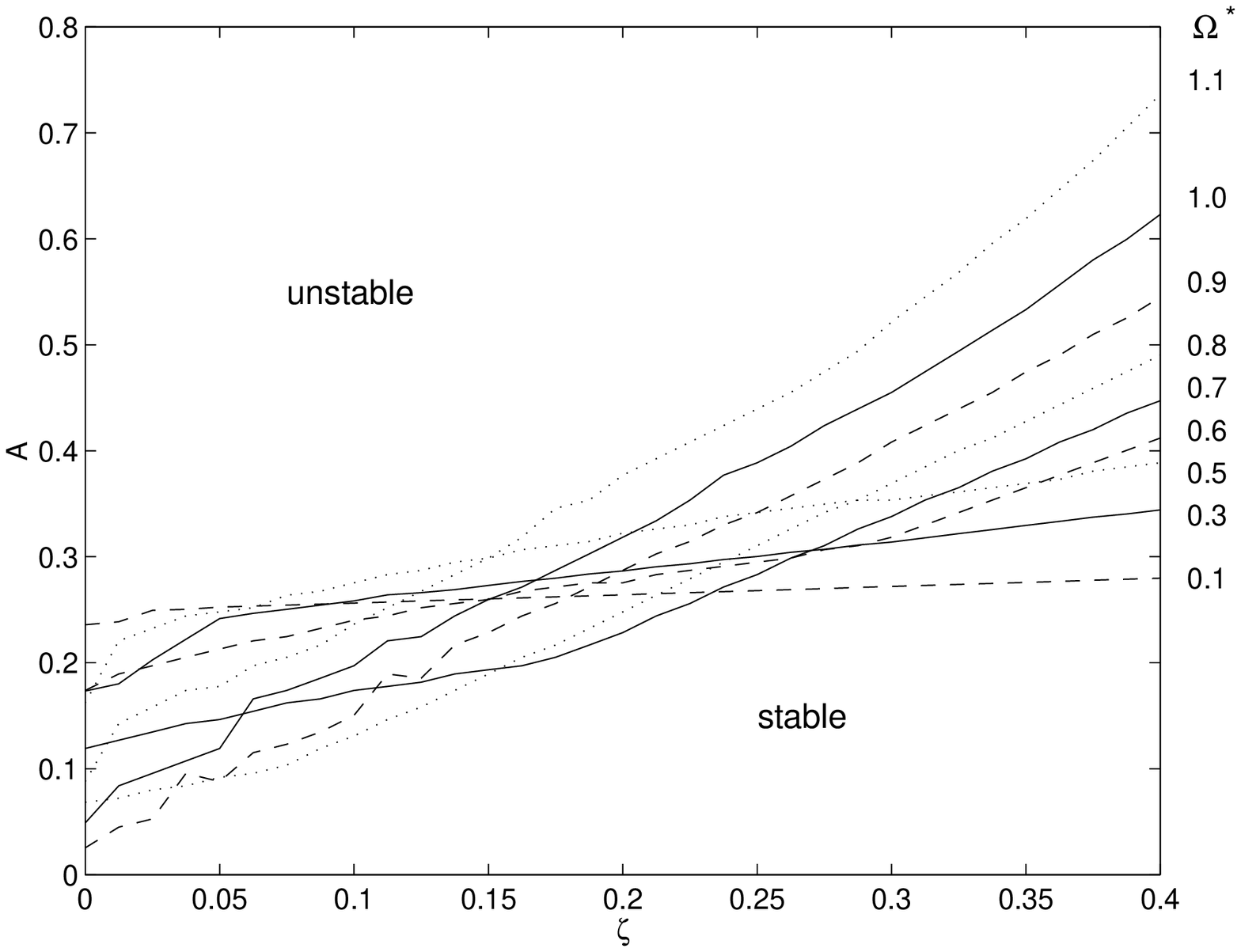,height=5in,width=5in}
\end{center}
\caption{}
\label{figure11}
\end{figure}

\begin{figure}[h!]
\begin{center}
\begin{minipage}{2.25in}
\epsfig{figure=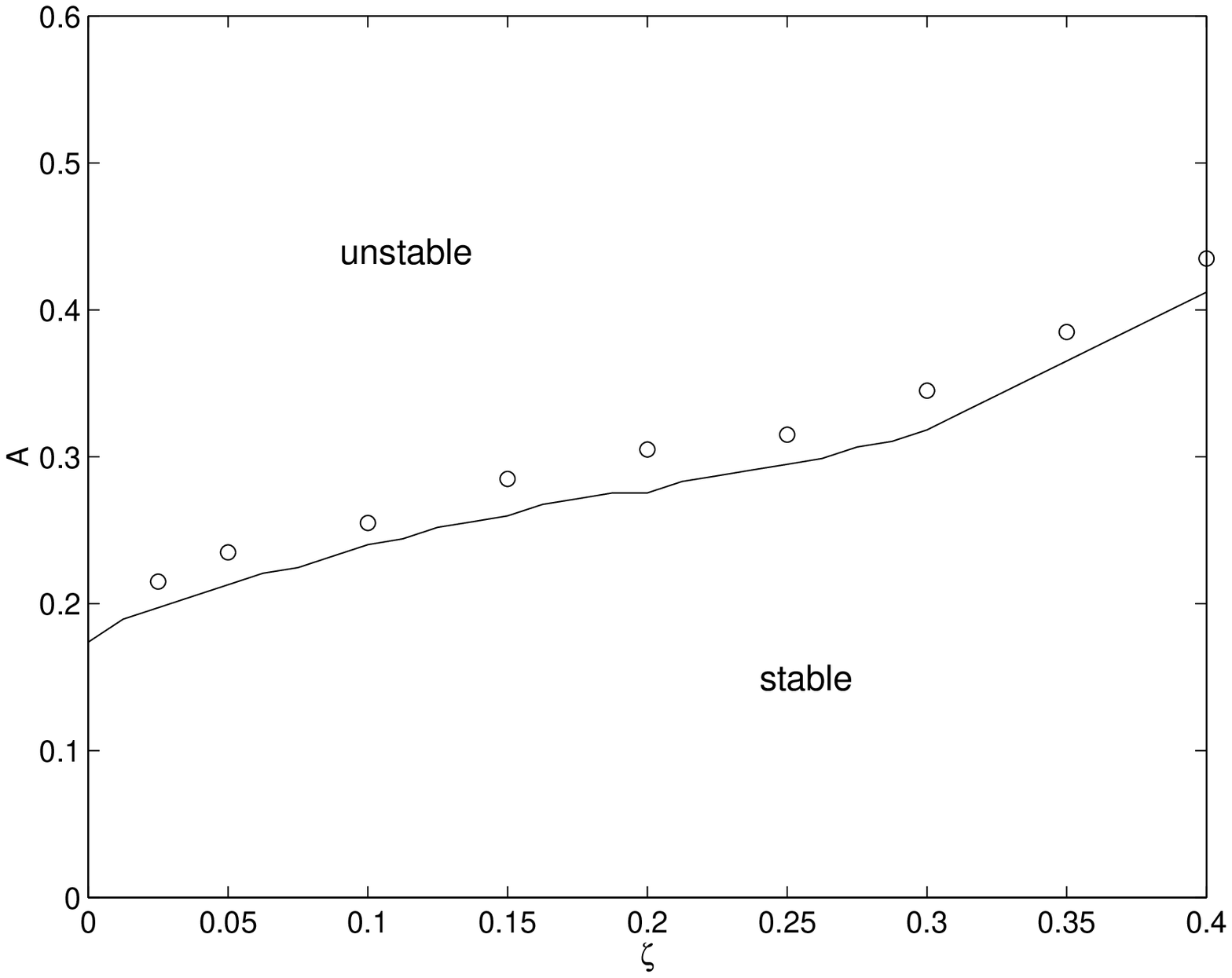,height=2.25in,width=2.25in}
\begin{center} (a) \end{center}
\end{minipage}
\hspace*{0.2in}
\begin{minipage}{2.25in}
\epsfig{figure=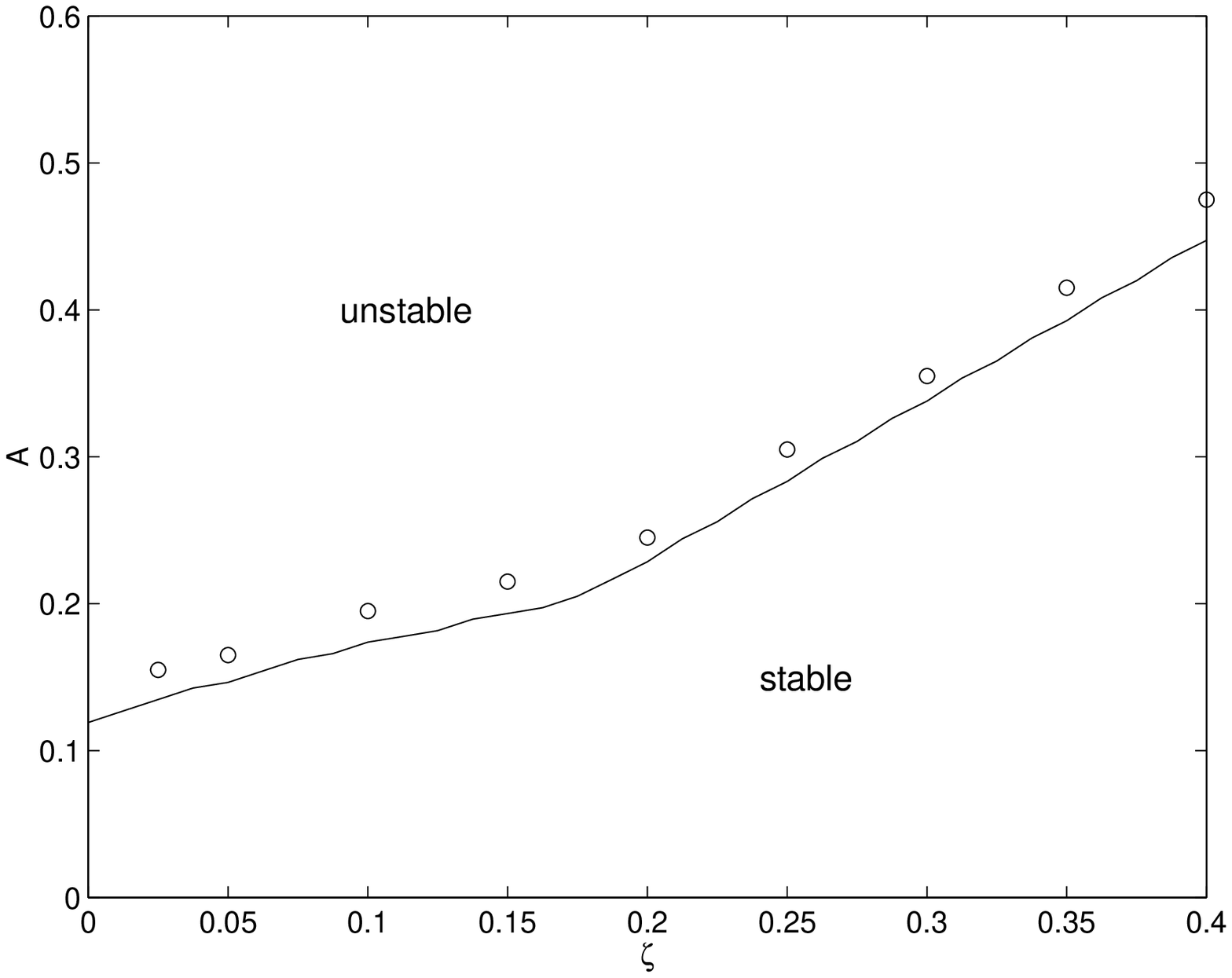,height=2.25in,width=2.25in}
\begin{center} (b) \end{center}
\end{minipage}
\end{center}
\vspace{0.2in}
\begin{center}
\begin{minipage}{2.25in}
\epsfig{figure=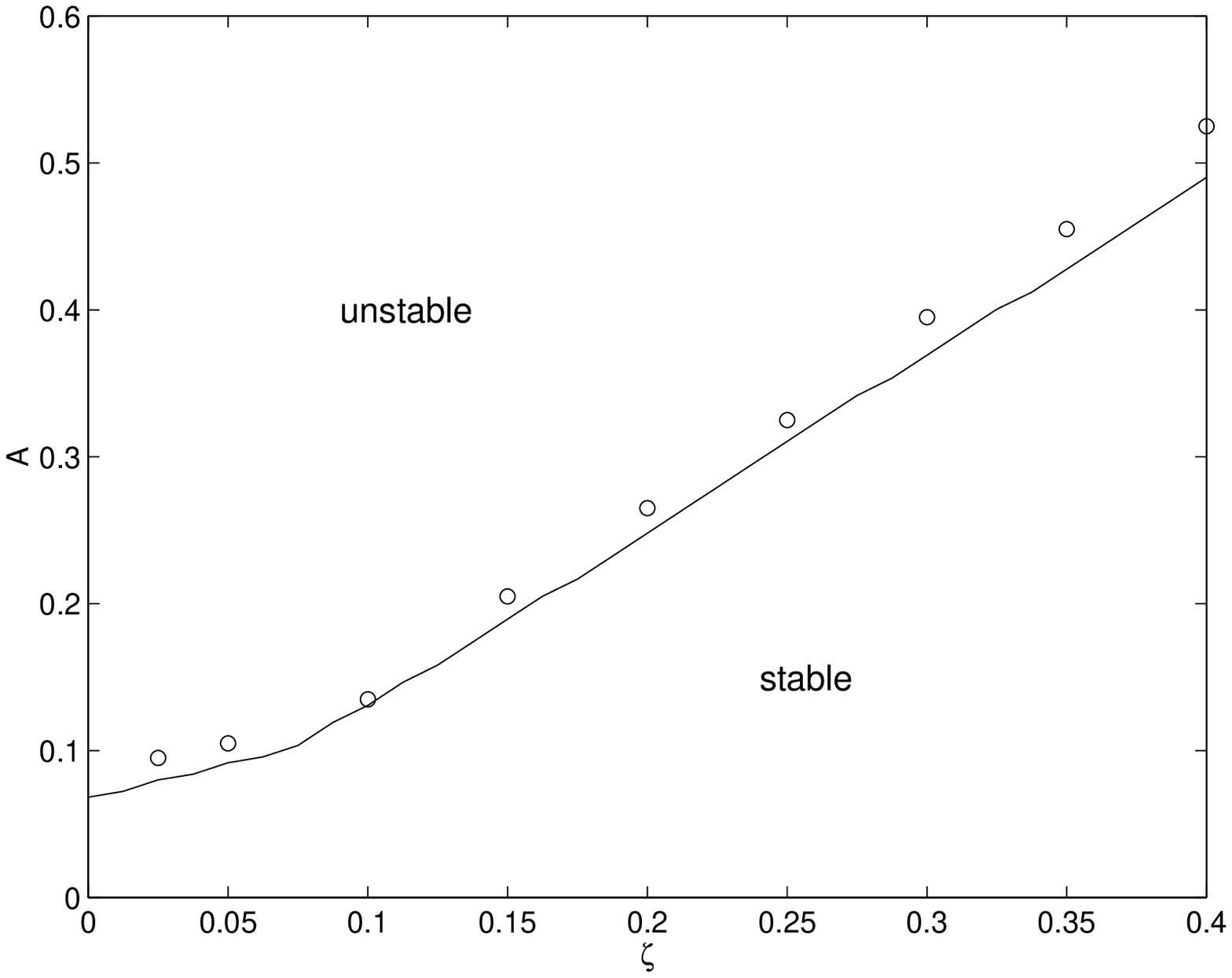,height=2.25in,width=2.25in}
\begin{center} (c) \end{center}
\end{minipage}
\hspace*{0.2in}
\begin{minipage}{2.25in}
\epsfig{figure=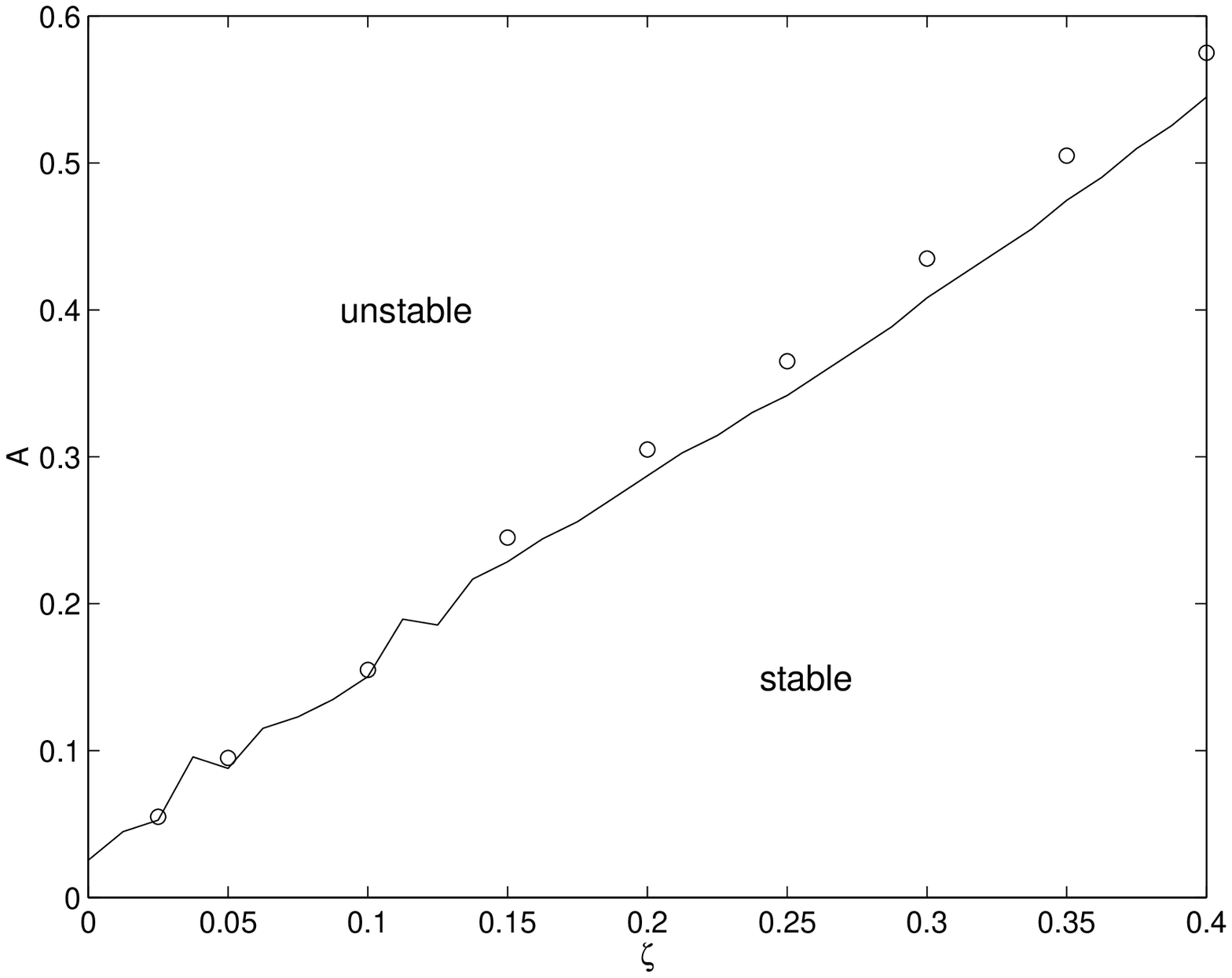,height=2.25in,width=2.25in}
\begin{center} (d) \end{center}
\end{minipage}
\end{center}
\caption{}
\label{figure12}
\end{figure}

\begin{figure}[h!]
\begin{center}
\begin{minipage}{2.7in}
\epsfig{figure=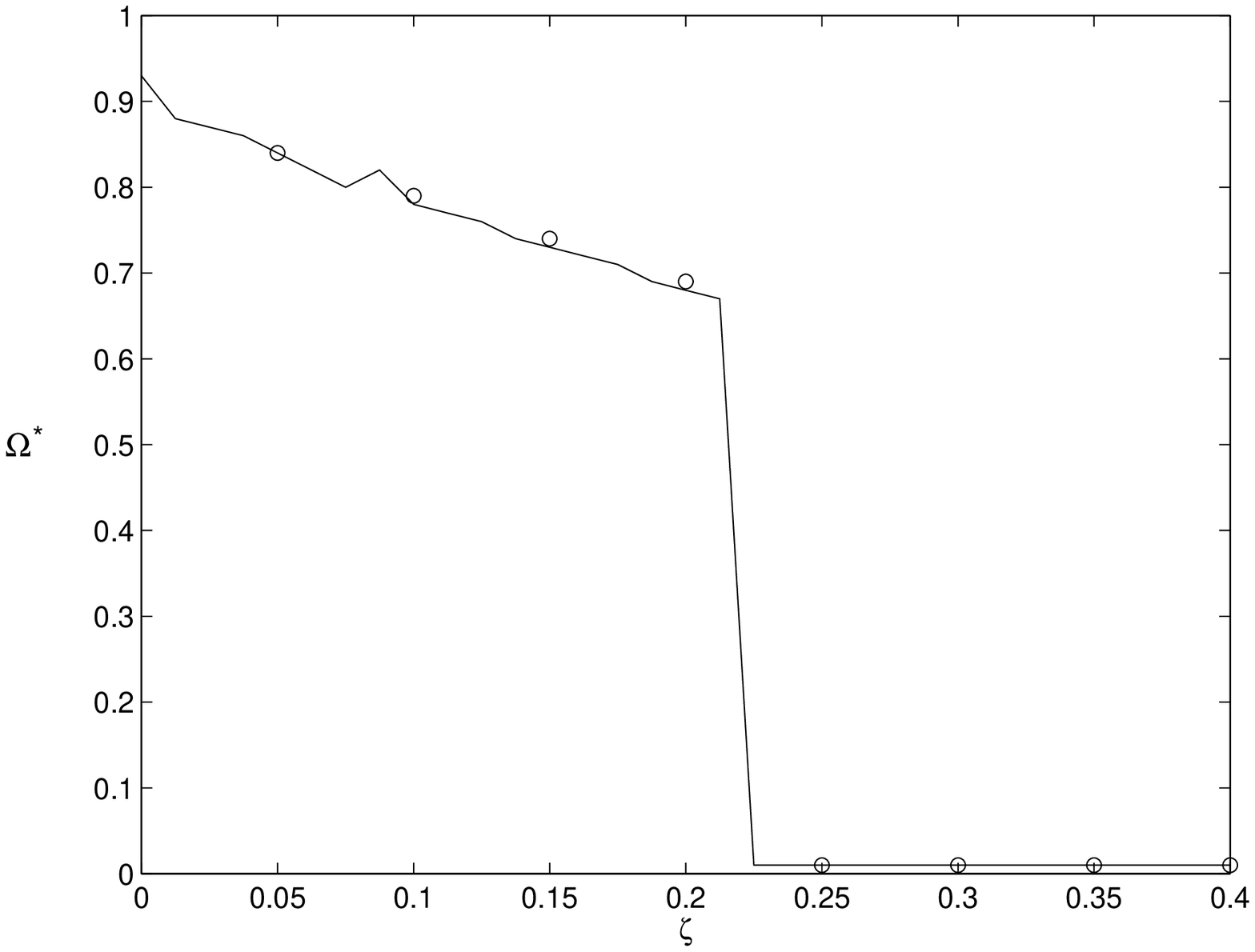,height=2.7in,width=2.7in}
\begin{center} (a) \end{center}
\end{minipage}
\hspace*{0.2in}
\begin{minipage}{2.7in}
\epsfig{figure=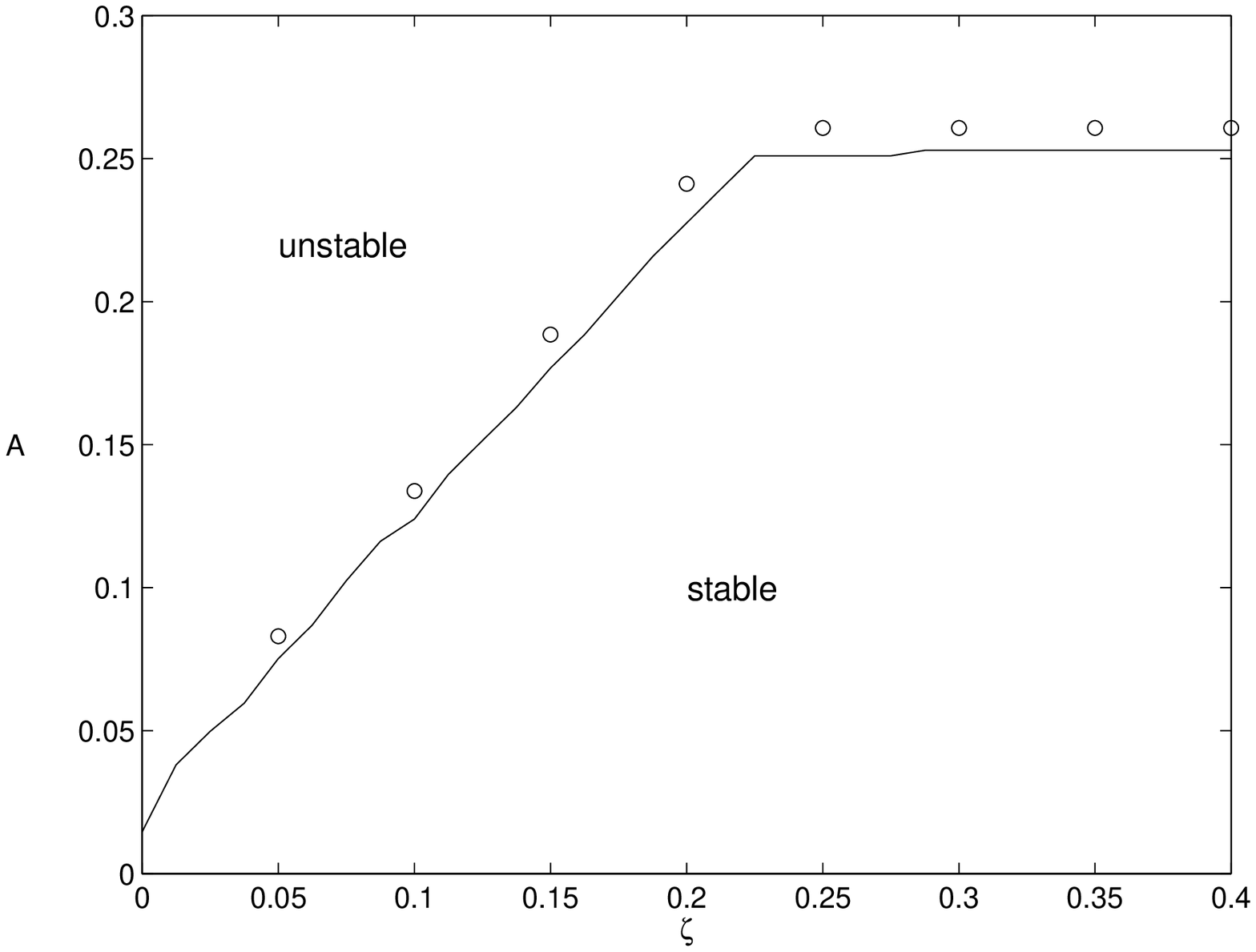,height=2.7in,width=2.7in}
\begin{center} (b) \end{center}
\end{minipage}
\end{center}
\caption{}
\label{figure13}
\end{figure}

\begin{figure}[h!]
\begin{center}
\begin{minipage}{2.7in}
\epsfig{figure=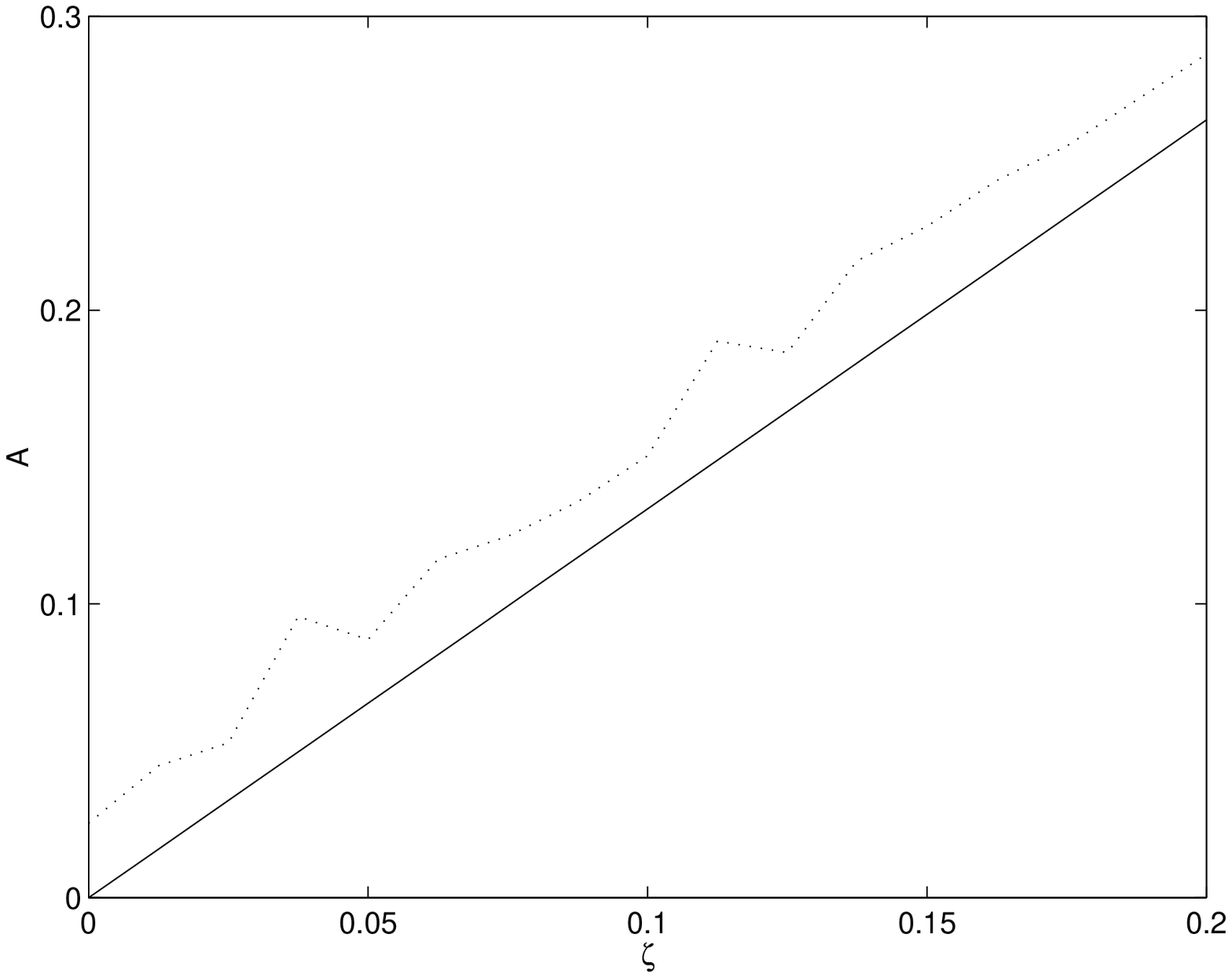,height=2.7in,width=2.7in}
\begin{center} (a) \end{center}
\end{minipage}
\hspace*{0.2in}
\begin{minipage}{2.7in}
\epsfig{figure=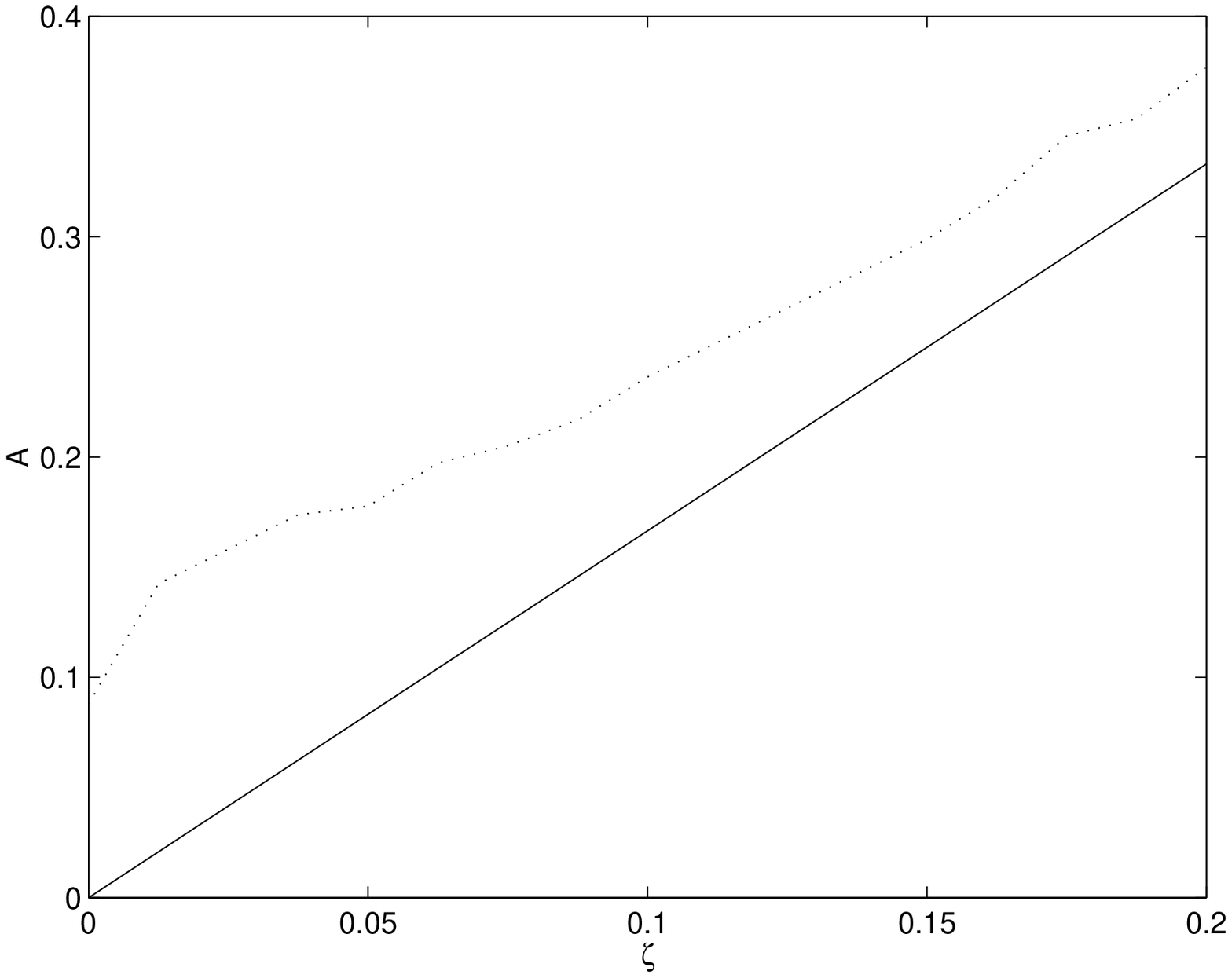,height=2.7in,width=2.7in}
\begin{center} (b) \end{center}
\end{minipage}
\end{center}
\caption{}
\label{figure14}
\end{figure}

\begin{figure}[h!]
\begin{center}
\epsfig{figure=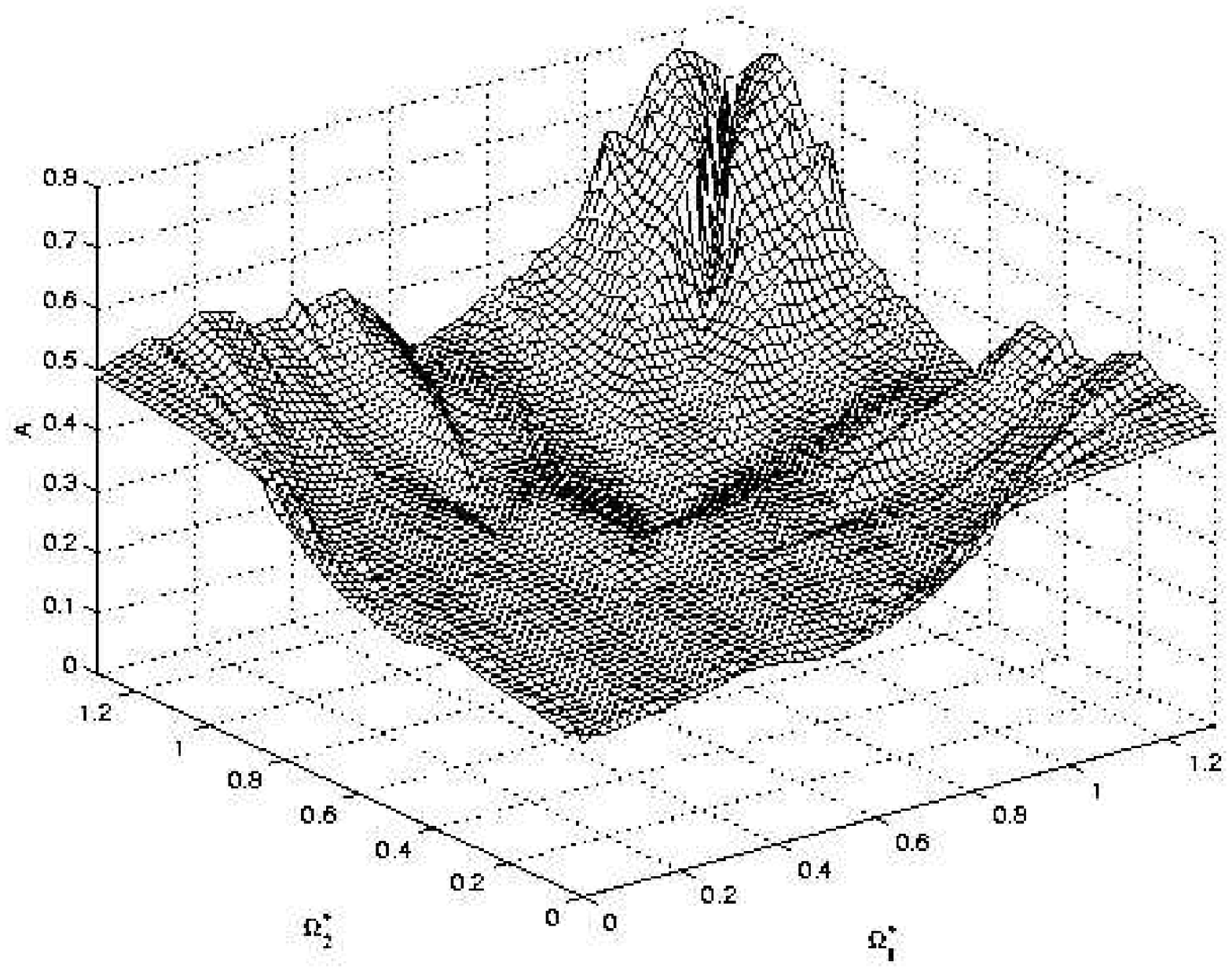,height=5in,width=5in}
\end{center}
\caption{}
\label{figure15}
\end{figure}

With the given physical parameters, and using the forcing pressure of
$p_A=1.5$ bar and $\Omega=2\pi \times 30 \times 10^{-6}$ s$^{-1}$, 
the parameters $\zeta$, $A$ and $\Omega^*$ are calculated from equation
(\ref{dist_lim_params}) to be: $\zeta=0.98$, $A=0.09$ and
$\Omega^*=0.16$.  Upon examining figure \ref{figure13}, at the
relatively large damping parameter $\zeta=0.98$ (beyond the range
originally considered) the minimum forcing threshold would appear to
correspond to the constant value of $A=0.25$.
Here we also note that this same forcing 
threshold is also observed with a range
of small $\Omega^*$,
including $\Omega^*=0.16$,
see figure \ref{figure11}.
In other words, the predicted threshold pressure for the active detector
to cause cavitation is $A=0.25$ which corresponds roughly to $p_A=4$
bar, whereas in the experiments the threshold was seen to be $A=0.09$
or $p_A=1.5$ bar. Despite the lack of quantitative agreement, the
theoretical predictions and the experiments do show the same trends.
Namely, in the absence of the 30-MHz detector, the pressure in the
liquid had to be reduced to $-15$ bar for cavitation to occur.  With
the high-frequency transducer turned on, however, cavitation occurred
at a minimum pressure of $-7-1.5=-8.5$ bar in the experiments and at
$-7-4=-11$ bar based on the theory. (We are adding the negative
pressure contribution from the two transducers to arrive at the final
minimum pressure). Thus, the presence of the second high-frequency
transducer does reduce the pressure threshold for cavitation in both
cases.

\vfill \break

%%%%%%%%%%%%%%%%%%%%%%%%%%% figure captions %%%%%%%%%%%%%%%%%%%%%%%%%%%

\clearpage
\begin{center}
{\Large \bf Figure Captions}
\end{center}

\fig{1}
Pressure in the liquid, $p_{L}$, versus bubble radius, $R$, as governed
by equation (\ref{liquid_pressure_iso}).

\fig{2}
Phase portraits for the distinguished limit equation
(\ref{dist_lim_eqn}).
In (a), $A=0$ and $\zeta = 0$.
The fixed point (0,0) is a center.
The fixed point (1,0) is a saddle.
In (b), $A=0$ and $\zeta =0.09$.
The fixed point (0,0) is a stable spiral.

\fig{3}
Poincar\'{e} section showing the unstable manifold of the saddle fixed point
(0.999769375,-0.024007197) for $A=0.048$ and $\zeta=0$.
Asymptotically, the saddle point is located at a distance
${\cal O}(A)$ from (1,0), specifically
($1-A^{2}/10 + {\cal O}(A^{4}), -A/2 - (13/200)A^{3} + {\cal O}(A^{5})$).
Invariant tori are shown inside a portion of the unstable manifold.
The center point has moved a large distance from $(0,0)$ in this case
due to the 1:1 resonance when $\Omega^{*}=1$.  And we note for
comparison that with nonresonant values of $\Omega^{*}$, Poincar\'{e} 
sections show that the center only moves an ${\cal O}(A)$ distance.
For example, when $\Omega^{*}=0.6,0.7,0.85$, the centers are located 
approximately at the points (0.009,0.045),(0.009,0.066) and
(0.029,0.163) respectively.

\fig{4}
Escape parameters for the trajectory of the origin (0,0).
For values of $A$,$\zeta$ above the regression line
the trajectory of the origin grows without bound.  Below this
line the trajectory of the origin remains bounded.
Least squares fit: $A=1.356 \zeta + 0.058$.

\fig{5}
Period doubling route to chaos in the distinguished limit equation
(\ref{dist_lim_eqn}).
For $\zeta=0.35$, the limit cycles undergo period doubling as
$A$ is increased.

\fig{6}
Bifurcation diagram ($\zeta = 0.375$).
Plotted is $\dot{x}$ versus $A$.  For each fixed value of $A$, the
origin (0,0) is integrated numerically and the value of $\dot{x}$
is plotted every $\Delta \tau = 2 \pi$.

\fig{7}
Bounded trajectories for the distinguished limit equation with
$\zeta=0.2$, $A=0.3$ and $\Omega^{*}=1.0$.
The dark region is the set of initial conditions whose trajectories
remain bounded; it is the basin of attraction of
the periodic orbit that exists 
in the period-doubling hierarchy
for this value of $A$.

\fig{8}
Simulations of the full Rayleigh-Plesset equation for four different
values of $\epsilon$.  Each open circle represents an $(A,\zeta)$ pair at
which the bubble first goes unstable.  Superimposed is the linear
regression line obtained from the simple criterion based upon the
distinguished limit equation.
In (a)--(d), $\epsilon = 0.01, 0.05, 0.1, 0.2$, respectively.

\clearpage

\fig{9}
Simulations of the full Rayleigh-Plesset equation for four different
values of $\epsilon$.  Each open circle represents a $(p_{A},R_{0})$ pair at
which the bubble first goes unstable.  Superimposed is the threshold
curve (\ref{stab_thresh}) obtained from the simple criterion based
upon the distinguished limit equation.
In (a)--(d), $\epsilon = 0.01, 0.05, 0.1, 0.2$, respectively.

\fig{10}
Radius versus time plots
$(\epsilon = 0.1, \, \zeta = 0.3, \, A=0.3,0.5,0.53)$.
\newline
Dimensional parameters:
$R_{0}=3.0 \, \mu$m,
$R_{crit} =3.2 \, \mu$m,
$p_{v}-p_{0}^{\infty} = 29.7 \,$kPa,
$\Omega = 0.7 \,$MHz.
\newline
Top: $p_{A}=141.6 \,$Pa.
Middle: $p_{A}=236.0 \,$Pa.
Bottom: $p_{A}=250.2 \,$Pa.

\fig{11}
Stability threshold curves for the origin trajectory of the
distinguished limit equation (\ref{dist_lim_eqn}) for many
different values of $\Omega^{*}$.  The values of $\Omega^{*}$
on the right hand border label the different threshold curves.

\fig{12}
Simulations of the full Rayleigh-Plesset equation for four different
values of $\Omega^{*}$.  In each plot $\epsilon=0.05$.
Each open circle represents an $A, \zeta$ pair at which the bubble first goes
unstable.  Superimposed is the threshold curve obtained from the
distinguished limit equation.
In (a)--(d), $\Omega^{*} = 0.6, 0.7, 0.8, 0.9$, respectively.

\fig{13}
In (a), the $\Omega^{*}$ that minimizes $A_{\rm esc}$ is
plotted versus $\zeta$.  In (b), the minimum value of
$A_{\rm esc}$ corresponding to the value of $\Omega^{*}$ in
(a) is plotted versus $\zeta$.  The open circles represent
Rayleigh-Plesset calculations of the minimum threshold with
$\epsilon = 0.05$.

\fig{14}
Comparisons of the stability curves of figure \ref{figure11} with
the Melnikov analysis for two values of $\Omega^{*}$.  The dotted
line is the stability curve obtained from the distinguished limit
equation.  The solid straight line is equation (\ref{homotan}).
In (a) and (b), $\Omega^{*} = 0.9, 1.1$, respectively.

\fig{15}
Stability threshold surface for the trajectory through the
origin obtained by integrating the distinguished limit equation with
quasiperiodic forcing.  This was obtained from simulations of 
(\ref{dist_lim_eqn}) with the forcing replaced by
$(A/2)(\sin (\Omega_{1}^{*} \tau ) + \sin (\Omega_{2}^{*} \tau ) )$,
and for $0 < \Omega_{1}^{*},\Omega_{2}^{*} < 1.3$.
The value of $\zeta$ was fixed at $\zeta=0.25$.
The points below the surface correspond to parameter values for which 
the trajectory of the origin remains bounded whereas those points above 
the surface are parameters which lead to an escape trajectory for the origin.
Qualitatively and quantitatively similar results
were obtained for $\zeta=0.15$ and $\zeta=0.35$.

%%%%%%%%%%%%%%%%%%%%%%%%%%% figures %%%%%%%%%%%%%%%%%%%%%%%%%%%

\figpage{1}

\figpage{2}

\figpage{3}

\figpage{4}

\figpage{5}

\figpage{6}

\figpage{7}

\figpage{8}

\figpage{9}

\figpage{10}

\figpage{11}

\figpage{12}

\figpage{13}

\figpage{14}

\figpage{15}

\end{document}